\documentclass[trackchanges,twocolumn]{aastex7}

\usepackage{amsmath}

\begin{document}

\title{Set the Night on FIRE: Building an Empirical Local Dark Matter Velocity Distribution}

\author[orcid=0000-0000-0000-0001,sname='Zhang']{Xiuyuan Zhang}
\affiliation{Department of Physics and Kavli Institute for Astrophysics and Space Research, MIT, Cambridge, MA 02139, USA}
\email[show]{xiuyuan@mit.edu}  

\author[orcid=0000-0000-0000-0002,sname='Thoyas']{Andreas Thoyas} 
\affiliation{Department of Physics, Northeastern University, Boston, MA 02115, USA}
\email{athoyas@mit.edu}

\author[0000-0003-2806-1414]{Lina Necib}
\affiliation{Department of Physics and Kavli Institute for Astrophysics and Space Research, MIT, Cambridge, MA 02139, USA}
\email{lnecib@mit.edu}  

\author[orcid=0000-0003-0603-8942]{Andrew Wetzel}
\affiliation{$^1$ Department of Physics \& Astronomy, University of California, Davis, CA 95616, USA}
\email{awetzel@ucdavis.edu}

\author[orcid=0000-0002-8354-7356]{Arpit Arora}
\affiliation{Department of Astronomy and DiRAC Institute, University of Washington, 3910 15th Ave NE, Seattle, WA, 98195, USA}
\email{aarora125@sas.upenn.edu}
 
\begin{abstract}
The majority of terrestrial direct detection experiments for Dark Matter (DM) rely on the Standard Halo Model (SHM), which assumes the local DM velocity distribution follows a Maxwell–Boltzmann distribution.
However, galaxy mergers can deposit DM that remains kinematically clustered today, inducing deviations from the smooth SHM prediction.
Previous studies have suggested that the local stellar velocity distribution may serve as a tracer for DM populations originating from the same progenitor systems.
In this work, we systematically investigate how merger mass and accretion time affect the correlation between local stellar and DM velocity distributions in Milky Way–like galaxies from the FIRE-2 simulations.
We find a strong correlation between traceable DM components and their stellar counterparts, with the tightest correspondence arising from lower-mass mergers accreted at earlier cosmic times.
For the remaining DM that lacks an identifiable stellar counterpart, which dominate the full DM fraction, we find that its velocity distribution is well described by a component-wise generalized Gaussian.
Combining these two ingredients, we reconstruct the full local DM velocity distribution.
This framework captures merger-induced features—such as co-rotation of accreted material with the galactic disk—that are entirely absent in the SHM.
Finally, we propagate uncertainties through the reconstruction and show that they are dominated by the stellar mass–halo mass relation, which is unlikely to improve substantially in the near term.
We therefore argue that this framework approaches the current limit of our ability to characterize the local DM velocity distribution.

\end{abstract}


\section{Introduction} 

Dark matter (DM)~\citep{1933AcHPh...6..110Z, 1980ApJ...238..471R, Bertone_2018} is a non-luminous component of matter whose total abundance is inferred to be approximately five times that of ordinary baryonic matter~\citep{Spergel_2003, 2003ApJS..148....1B}. It has remained a central open problem in both particle physics and astrophysics for several decades. Observations indicate that every galaxy is embedded in an extended DM halo that dominates its mass budget~\citep{1973ApJ...186..467O, einasto2004darkmatterearlyconsiderations, Bertone_2018}. Thus, one major strategy for probing DM in terrestrial experiments is direct detection, which aims to observe rare scattering events between DM particles and Standard Model targets~\citep{PhysRevD.30.2295, PhysRevD.31.3059, PhysRevD.33.3495, RevModPhys.85.1561}. All of these searches rely critically on the assumed flux of DM passing through the Earth, which is determined by the local DM density and velocity distribution.

The most commonly adopted assumption for the DM velocity distribution is the Standard Halo Model (SHM)~\citep{1986PhRvD..33.3495D, 1996APh.....6...87L}, which describes the Galactic DM halo as an isothermal, spherically symmetric system in equilibrium. In this framework, the local DM velocities follows a Maxwell–Boltzmann distribution,\footnote{The isothermal solution does not formally include a cutoff; in practice, a cutoff at the Galactic escape speed is imposed, typically $v_{\rm{esc}}\sim500-600~\rm{km/s}$~\citep[see e.g.][]{Smith_2007, Monari_2018, 2021A&A...649A.136K,2022ApJ...926..189N,2024ApJ...972...70R}.} and the spherical Jeans' equation relates the one-dimensional DM velocity dispersion to the circular velocity via $\sigma_{DM} = v_c/\sqrt{2}$, where a flat rotation curve with $v_c \sim 220~\rm{km/s}$ is assumed~\citep{1986MNRAS.221.1023K, 2016ARA&A..54..529B, 2019ApJ...871..120E}.

The isothermal profile underlying the SHM assumes $\rho \propto r^{-2}$, with $r$ the spherical distance from the center of galaxy, producing this flat rotation curve -- a simplification that deviates from the Milky Way's actual mass distribution. While this deviation is small at the solar position, more generally, the local velocity distribution depends on the halo’s density slope, any radial variation in the velocity dispersion $\sigma$, and the velocity anisotropy. Furthermore, the Milky Way's halo is not expected to be in equilibrium due to its hierarchical assembly history and ongoing merger events, so the true DM velocity distribution is expected to depart from the Maxwell-Boltzmann form assumed in the SHM. 

Several groups have used cosmological simulations to quantify these departures and their consequences for direct detection. Using high-resolution Milky Way–like halos, \cite{Vogelsberger_2009} found that nuclear recoil rates can differ from the SHM prediction by up to 10\%. \cite{Mao_2013, Mao_2014} characterized the halo-to-halo scatter in the local velocity distribution, showing that it can appreciably shift exclusion limits and preferred parameter regions. Analyses based on Via Lactea–type halos~\citep{2007ApJ...667..859D, 2008Natur.454..735D} explicitly compared simulation-derived recoil spectra against SHM predictions, finding that light and inelastic DM scenarios are particularly sensitive to structure in the high-velocity tail~\citep{Friedland_2013}. More recently, hydrodynamical simulations incorporating baryons have revealed that disk formation and baryonic contraction reshape the local DM phase-space distribution, further modifying direct-detection interpretations~\cite{Bozorgnia_2017}. Taken together, these studies establish that the assumed velocity distribution leads to systematic uncertainties in direct-detection analyses.

Recent developments in observational astrometry have shown that the Milky Way (MW) has undergone multiple mergers throughout its history~\citep[see e.g.,][for a review]{2020ARA&A..58..205H}. The most prominent merger is the Gaia–Sausage–Enceladus (GSE)~\citep{2018MNRAS.478..611B,2018Natur.563...85H}; 
the GSE is assumed to have been accreted to the MW about 10 billion years ago, and was the result of a 1:3 merger~\citep{johnson2025thatsretrogaiasausageenceladusmerger} that contributed around one-fourth of the total mass of the MW and is likely responsible for thickening the MW's disk~\citep{2018Natur.563...85H, 2020A&A...642L..18K}. Other mergers include the currently stripped Sagittarius Dwarf Spheroidal Galaxy (Sgr dSph)~\citep{1994Natur.370..194I, Ibata_1995},  
as well as the infalling satellites like the Large and Small Magellanic Cloud (LMC and SMC)~\citep{2006ApJ...638..772K, Besla_2007} that also potentially affects the local DM velocity distribution~\citep{Besla_2019}. 

It was then proposed in ~\cite{2018PhRvL.120d1102H} that the stellar velocity distribution can serve as a proxy for the full DM velocity distribution. Subsequently, using two Milky Way–like galaxies from the FIRE-2 simulations~\citep{2016ApJ...827L..23W,Hopkins_2018}, \cite{Necib_2019} demonstrated a correlation between the velocity distributions of different stellar populations and distinct DM components in the solar neighborhood. In particular, DM and stars associated with early mergers (at redshift $z \gtrsim 3$) are dynamically relaxed and therefore exhibit a strong correspondence. By contrast, material from more recent mergers is only partially phase-mixed, leaving behind debris flows and cold streams. While the stellar–DM correspondence remains adequate for debris-flow components, it becomes significantly less reliable for cold streams (with infall times $z<1$) due to their pronounced spatial inhomogeneity. Consequently, a more sophisticated treatment is required to reconstruct the underlying DM velocity distribution from stellar tracers. Subsequent work in ~\cite{shpigel2025buildempiricalspeeddistribution} showed that, in the presence of a single late-time merger, it is nevertheless possible to recover the DM speed distribution accurately by incorporating a simple boost factor into the stellar-based reconstruction.

In this work, we investigate seven \texttt{m12} FIRE galaxies, comprising a total of 24 major mergers, to establish a concrete and self-consistent reconstruction of the local DM velocity distribution. We begin by decomposing the local DM population into two major components: DM accreted through mergers, and a smoothly accreted diffuse component. The merger component is then further divided into resolved and unresolved merger components. The resolved component is discussed in Sec.~\ref{sec:Correlation}. Ranking by their contributions to the local accreted stellar mass, the top four contributors are designated as traceable mergers (Sec.~\ref{sec:Traceable}), while the remaining is the untraceable merger component (Sec.~\ref{subsec:Untractable mergers}), which is further divided into young and old untraceable mergers by their accretion time. The unresolved component and untraceable mergers are then combined together with the diffuse component to define a single relaxed component. Thus the local velocity component can be structured as the traceable mergers and the relax component,
\begin{equation}\label{full_distribution}
    f_{\rm{DM}}(v) = \sum_{i=1}^4 c_i f_i(v) + (1-\sum_{i=1}^4 c_i) f_{\rm{relaxed}}(v),
\end{equation}
where $c_i$ is the mass fraction of local DM from the $i$-th traceable merger, $f_i$ is the corresponding velocity distribution, and $f_{\rm{relaxed}}$ is the velocity distribution of the relaxed component.

We establish a connection between the stellar and DM velocity distributions originating from the same merger, and characterize how this connection varies with accretion time. We model the diffuse component as a generalized Gaussian function and demonstrate, using simulations, that the total local DM velocity distribution can be robustly reconstructed. We assess both the sensitivity of this reconstruction to the numerical resolution of the simulations and the main sources of error that would arise in an application to the Milky Way. 

Alternatively, we also demonstrate that the generalized Gaussian by itself can be directly used as an approximation of the full velocity distribution. In terms of speed distribution, it is a simple yet consistent improvement over SHM. However, it might fail to predict the azimuthal velocity distribution in the presence of co-rotation. 

This paper is organized as follows. In Sec.~\ref{sec:sim}, we introduce the FIRE simulation suite: we summarize the key properties of the various \texttt{m12} galaxies in Sec.~\ref{subsec:FIRE}, and describe our methodology for associating DM and stellar particles with individual mergers and tracing them back in time in Sec.~\ref{subsec:BuildMerger}. In Sec.~\ref{sec:Correlation}, we start with a discussion of the merger component and examine the properties of DM and stellar components associated with major mergers. 
We begin by introducing the statistical framework in Sec.~\ref{subsec:EMD}, analyze the dependence of stellar–DM correlations on merger accretion times in Sec.~\ref{subsec:AH}, and present a robust comparison of the full three-dimensional velocity distributions in Sec.~\ref{subsec:3d}. We further investigate the use of boost factors to better align stellar and DM velocity distributions in Sec.~\ref{subsec:Matching}. 
In Sec.~\ref{sec:Diffuse}, we turn to the diffuse DM component, showing that it dominates the local DM budget and is well described by a generalized Gaussian velocity distribution. Building on this, Sec.~\ref{sec:Reconstruction} outlines our method for reconstructing the local DM speed distribution from stellar kinematic data alone. Additionally, we demonstrate that the generalized Standard Halo Model (gSHM) as a standalone DM distribution model provides a consistent improvement over the traditional SHM. We discuss the limitations and uncertainties of our approach in Sec.~\ref{subsec:Uncertainties}, and assess the robustness of our results across simulation resolutions in Sec.~\ref{subsec:Resolutions}. We conclude in Sec.~\ref{sec:Conclusion}.

\section{Simulations} \label{sec:sim}

\subsection{FIRE-2 Simulation} \label{subsec:FIRE}

\begin{figure*}[t]
    \centering
    \includegraphics[width=0.7\linewidth]{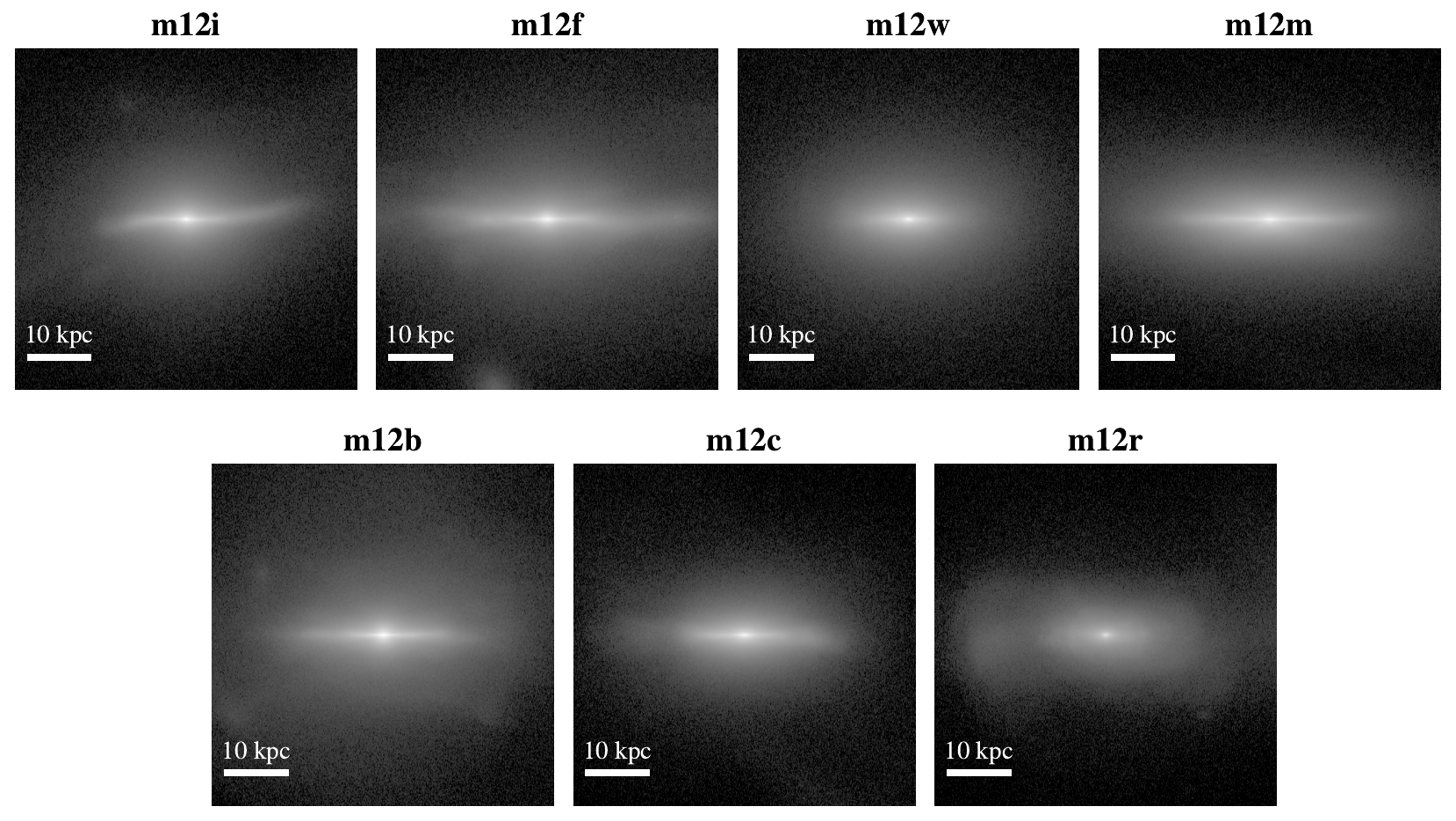}
    \caption{\label{fig:edgeon} Edge-on stellar surface density maps at $z=0$ for the seven Milky Way--mass simulations: \texttt{m12i}, \texttt{m12f}, \texttt{m12w}, \texttt{m12m}, \texttt{m12b}, \texttt{m12c}, and \texttt{m12r}. Stellar particles are shown within $|x| \le 30$ kpc and $|z| \le 30$ kpc in the principal-axis frame of the host galaxy. White regions correspond to higher stellar surface density. The figure highlights morphological diversity in disk structure among the simulated Milky Way analogs.
    }
\end{figure*}

In this work, we analyze the full merger history of the \textit{Latte} suite of cosmological zoom-in hydrodynamic simulations with FIRE-2 physics—namely \texttt{m12i}, \texttt{m12f}, \texttt{m12w}, \texttt{m12m}, \texttt{m12b}, \texttt{m12c}, and \texttt{m12r}—together with their corresponding lower-resolution realizations across 600 snapshots from redshifts $z=0-99$. We also include the higher-resolution “Triple Latte” rerun of \texttt{m12i} to be introduced in Wetzel et al., in prep.

All runs employ the Feedback In Realistic Environments (FIRE)-2 galaxy-formation model implemented in GIZMO with the meshless finite-mass (MFM) hydrodynamics solver~\citep{2014MNRAS.445..581H, Hopkins_2015, 2016ApJ...827L..23W, Hopkins_2018}. Gas follows radiative cooling and heating from $\sim10$–$10^{10}\mathrm{K}$, including primordial and metal-line processes, Compton and free–free emission, molecular/dust cooling at low temperatures, and photoheating by a redshift-dependent UV background (with local self-shielding)~\citep{2009ApJ...703.1416F}. Star formation is restricted to dense, locally self-gravitating gas (typical criterion $n_{\rm H}\ge10^3,\mathrm{cm^{-3}}$) and once a gas cell meets the criteria for star formation, it converts to a star particle on a local free-fall time (with the global efficiency regulated by feedback). Each star particle represents a single-age stellar population drawn from a Kroupa IMF~\citep{2001MNRAS.322..231K}. The feedback model is time-resolved and includes Type II and Ia supernovae, OB and AGB stellar winds, photoionization and photoelectric heating, and radiation pressure. Mass, metals, energy, and momentum are returned to neighboring gas elements in a manifestly conservative manner, tracking multiple metal species with subgrid turbulent metal diffusion.

The baryonic and DM mass resolutions and adopted cosmologies for the \textit{Latte} suite are summarized in Table~\ref{tab:sims}. Each halo is selected as a MW analog with virial mass $\sim10^{12}M_\odot$ and is isolated, with no comparably massive halo within five virial radii.\footnote{We define the virial radius $R_{200}$ as the radius enclosing an average density $200\rho_{\mathrm{c}}$, where $\rho_{\mathrm{c}}=3H(t)^2/(8\pi G)$, and the virial mass as the mass contained within the virial radius. } The simulations adopt two standard cosmologies: `A' denotes the AGORA cosmology ($\Omega_{\rm m}=0.272$, $\Omega_{\Lambda}=0.728$, $\Omega_{\rm b}=0.0455$, $h=0.702$, $\sigma_8=0.807$, $n_s=0.961$) and `P' denotes the Planck cosmology ($\Omega_{\rm m}=0.31$, $\Omega_{\Lambda}=0.69$, $\Omega_{\rm b}=0.0458$, $h=0.68$, $\sigma_8=0.82$, $n_s=0.97$). Gravitational softening (Plummer-equivalent) are $4~\mathrm{pc}$ for stars and $40~\mathrm{pc}$ for DM; the minimum adaptive gas softening is $1~\mathrm{pc}$. The lower-resolution counterparts use particle masses coarser by a factor of $\sim8$ and softening larger by $\sim2$, whereas Triple Latte attains $\sim8$ times finer mass resolution (i.e., particle masses smaller by $\sim8$) with softening smaller by $\sim2$; in all cases the cosmology matches that of the corresponding parent run in~\cite{2016ApJ...827L..23W}. 

Across the \texttt{m12} suite, we find that accreted stars preferentially occupy prograde orbits over a wide range of metallicities. The primary exception is \texttt{m12i}, which exhibits a pronounced prograde bias only at higher metallicity~\citep{2021MNRAS.505..921S}. This behavior is broadly in agreement with the observation of the MW, which is largely dependent on the major merger history. 

  \begin{table}[t]
  \centering
    \begin{tabular}{|m{0.1\linewidth}|m{0.2\linewidth}|m{0.15\linewidth}|m{0.15\linewidth}|m{0.2\linewidth}|}
  \hline
   \centering Name & \centering $\rm{M_{halo}}[M_{\odot}]$& \centering $\rm{m_{star}}[M_{\odot}]$ & \centering $\rm{m_{dm}}[M_{\odot}]$& \centering Cosmology\tabularnewline
  \hline
  \centering \texttt{m12w} & \centering $1.08\times 10^{12}$ & \centering 7100 & \centering 39000 & \centering P \tabularnewline
  \hline
  \centering \texttt{m12r} & \centering $1.10\times 10^{12}$ & \centering 7100 & \centering 39000 & \centering P \tabularnewline
    \hline
  \centering \texttt{m12i} & \centering $1.18\times 10^{12}$ & \centering 7100 & \centering 35000 & \centering A \tabularnewline
    \hline
  \centering \texttt{m12c} & \centering $1.35\times 10^{12}$ & \centering 7100 & \centering 35000 & \centering A \tabularnewline
    \hline
  \centering \texttt{m12b} & \centering $1.43\times 10^{12}$ & \centering 7100 & \centering 35000 & \centering A \tabularnewline
    \hline
  \centering \texttt{m12m} & \centering $1.58\times 10^{12}$ & \centering 7100 & \centering 35000 & \centering A \tabularnewline
    \hline
  \centering \texttt{m12f} & \centering $1.71\times 10^{12}$ & \centering 7100 & \centering 35000 & \centering A \tabularnewline
     \hline    
  \end{tabular}
    \caption{\label{tab:sims}Properties of the \texttt{m12} simulations in this paper, including the halo mass $\rm{M}_{\rm{halo}}$, stellar and DM particle mass, $\rm{m_{star}}$ and $\rm{m_{dm}}$, and Cosmology that specifies energy density ratios, Hubble constants, $\sigma_8$ and $n_s$. `P' stands for Planck and `A' stands for AGORA. \texttt{m12r} is not further included for analysis due to a late time equal mass merger that fully disrupted the stellar disk making it an inappropriate MW analog. }
  \end{table}

\subsubsection{m12i}

Among these simulations, \texttt{m12i} is the first high-resolution MW halo to be extensively analyzed and remains a particularly well-studied analog: it has a total stellar mass $M_\star \simeq 5.5\times10^{10}M_\odot$, consistent with the MW’s $(5\pm1)\times10^{10}M_\odot$~\citep{2015ApJ...806...96L, McMillan_2016}, and at $z=0$ hosts a thin, extended stellar disk with a prominent spiral pattern and bar~\citep{2014MNRAS.445..581H, Garrison_Kimmel_2018, ansar2024barformationdestructionfire2}. It also reproduces reasonably the MW satellite luminosity function and spatial distribution~\citep{2019MNRAS.487.1380G, 2020MNRAS.491.1471S}, and—critically—exhibits a quiescent recent merger history (post-$z\sim1$), including a somewhat plausible analog of the Gaia-Sausage/Enceladus event (a relatively massive, radial merger at $z\sim1.2$ [see Table.~\ref{tab:m12_mergers}]). However, its strongly warped, flared outer stellar disk—the outer disk whose mid-plane bends and becomes noticeably misaligned with the inner disk and with the halo symmetry axes~\citep{2020ApJS..246....6S, baptista2022orientationsdmhalosfire2}, more so than in many of the other \texttt{m12} runs as shown in Fig.~\ref{fig:edgeon}, which makes it an imperfect match to the MW in detail. \footnote{Selection criteria for GSE-analogs are discussed in Appendix \ref{subsec:GSE}.}

\subsubsection{m12f}

Compared to \texttt{m12i}, \texttt{m12f} sits closer to an intermediate “MW/M31-like” regime: it is more massive, with a higher stellar mass and a rotationally supported disk that experiences short-lived bar episodes at late times~\citep{ansar2024barformationdestructionfire2, quinn2025spiralstructurepropertiesdynamics}. There is no strongly misaligned or polar inner disk as in \texttt{m12i}, but the disk is more massive, thicker, and has a higher–surface-density than the Milky Way’s~\citep{2024MNRAS.527.6926M}. 

In terms of accretion history, \texttt{m12f} undergoes two major mergers at redshifts $z \sim 1$ and $z \sim 0.1$. The earlier event is broadly consistent with the estimated infall time of GSE, while the later one occurs too recently to serve as a GSE analog. The DM halo masses at infall for these mergers are $\sim8.2\times10^{10}M_{\odot}$ and $\sim1.5\times10^{11}M_{\odot}$, respectively—both within the expected range for GSE-like progenitors. Each merger exhibits a stellar velocity anisotropy of $\beta \sim 0.7$ at $z=0$, indicating moderately radial orbits, albeit slightly less radial than the canonical GSE value.

However, when considering their stellar contributions to the solar neighborhood, both mergers deviate from standard GSE expectations. The $z \sim 1$ merger contributes approximately $18.5\%$ of the local accreted stellar population, which is too small to be considered a dominant GSE-like progenitor. In contrast, the more recent $z \sim 0.1$ merger accounts for around $44\%$ of local accreted stars, placing it near the lower end of observational estimates for the GSE contribution, but its very late accretion time makes it inconsistent with the expected GSE epoch. Together, these results suggest that while both mergers share individual characteristics with GSE, neither serves as a definitive GSE analog in all respects.

\subsubsection{m12b}

\texttt{m12b} is a comparatively quiescent Milky Way analog, with a relatively quiet late-time assembly history. Apart from a single massive  satellite accreted at $z\sim0.25$ which accounts for $72\%$ of the accreted material, all major mergers occur at $z\ge2$. By $z=0$, it hosts a barred, rotationally supported disk with roughly MW-like mass and size, though again slightly more massive, more compact, and dynamically thicker than the real Galaxy~\citep{Hopkins_2019, 2024MNRAS.527.6926M}. The disk is not as dramatically warped as in \texttt{m12i}; the outer-disk bending and flaring appear modest~\citep{Barry_2023, 2024ApJ...974..286A}. The late accreted satellite overwhelmingly dominates the accreted stellar component and is on a moderately radial orbit, with $\beta \sim 0.6$, though it is less radially biased than a canonical GSE event. Nevertheless, its accretion at $z \sim 0.25$ is likely too late for it to serve as a clean GSE analog, despite its dominant contribution and radial kinematics.

\subsubsection{m12w}

\texttt{m12w} is a MW-mass halo with a noticeably more active late-time assembly history than systems such as \texttt{m12b}. At $z=0$, it remains dynamically hot, with a relatively thick, disturbed stellar disk and a bursty late-time star-formation history, making it a poor morphological match to the observed Milky Way~\citep{Yu_2021, 2021ApJ...908L..31O, Arora_2022, Yu_2023}. One of its late mergers is a reasonably GSE-like, radial $1:8$ event at an appropriate lookback time and mass scale, and it dominates the accreted stellar population in the solar neighborhood. However, this interpretation is not clean: another merger of comparable mass occurs around a similar epoch even though on a more tangential orbit, contributing relatively little to the local stellar content, and the radial merger event itself occurs at $z\sim0.5$, somewhat later than typically expected for a canonical GSE analog. The late merger sequence is biased toward prograde orbits~\citep{Deason_2011, 2021MNRAS.505..921S}. As a result, the accreted material carries coherently aligned angular momentum that is deposited into the inner halo. This torques the pre-existing DM and produces substantial co-rotation of the dark halo with the stellar disk at $z=0$;~\citep{garavitocamargo2023corotationmilkywaysatellites, Arora_2025} the effect primarily reflects the angular momentum of the accreted DM rather than simple in-situ stellar heating~\citep{Read_2008, 2016MNRAS.460.4466Z, 2023ApJ...958...44B}. In practice, \texttt{m12w} is therefore useful as a laboratory for studying GSE-like satellites in an MW-mass system, but imperfect due to halo–disk torques, and active late-time accretion.

\subsubsection{m12m}

\texttt{m12m} sits at the massive, M31-like end of the suite: it resides in a halo of $M_{200}\sim1.58\times 10^{12}M_{\odot}$ and hosts $\sim 10^{11}M_{\odot}$ in stars, making it significantly more massive and more richly populated with satellites than the Milky Way. Its early assembly is dominated by two comparably massive components—the main branch and a gas-rich building block on a largely prograde orbit—which merge to form a proto-galaxy whose old, metal-poor stars remain in a very thick, strongly rotating, flattened configuration, co-rotating with the eventual disk at $v_{\phi}\sim100\rm{km/s}$~\citep{2021MNRAS.505..921S, horta2023protogalaxymilkywaymasshaloes}. As a result, \texttt{m12m} develops a pronounced prograde bias in its ancient stellar populations rather than a GSE-like, kinematically hot, nearly non-rotating inner halo. Morphologically, \texttt{m12m} forms a large, dynamically hot stellar disc and a strong bar only at late times (bar formation at $z\sim0.2$)~\citep{Garrison_Kimmel_2018, 2024MNRAS.527.6926M}, with bar buckling producing an X-shaped bulge that is heavily contaminated by stars younger than $\sim10$ Gyr—again in tension with the predominantly old ages inferred for the Galactic bulge. Taken together, \texttt{m12m} is interesting as a massive, bar-dominated MW-mass system with an MW-like ensemble early accretion history, but it is not a particularly clean Milky Way analog: it is too massive, its bar forms too late, its old stars rotate too strongly, and its dominant early building block does not manifest as a neat, GSE-like radial “sausage” in the inner halo.

\subsubsection{m12c}

\texttt{m12c} sits quite close to the Milky Way in bulk properties: its halo mass $M_{200}\sim1.35\times 10^{12}M_{\odot}$ and stellar mass $\sim5-6\times 10^{10}M_{\odot}$ are broadly MW-like, and by $z=0$ it hosts an extended, fairly settled stellar disk that is somewhat older, redder, and more gas-poor than the real Galaxy’s, with at most a weak or transient bar~\citep{Garrison_Kimmel_2018, Yu_2023, ansar2024barformationdestructionfire2}. Its early assembly, however, is not dominated by a single GSE-like event, but by a couple of relatively heavy mergers of comparable mass that fall in rapid succession around $z\sim1-1.5$. As a result, the inner accreted halo is built from several similarly important progenitors rather than exhibiting a clean, single radial “sausage,” and there is no obvious GSE analog. After $z\sim1$, \texttt{m12c}'s accretion history is comparatively quiet. Thus, while \texttt{m12c} is reasonably MW-like in present-day mass and the existence of a fairly undisturbed $z=0$ disk, it is a poor analog for the detailed MW assembly history. 

\subsubsection{m12r}

Finally, \texttt{m12r} is disfavored as a Milky Way analog: it experiences a very massive merger at $z\sim0.09$ that effectively destroys the stellar disk at late times~\citep{ma2025twophaseformationgalaxiescoevolution}, as can be seen in Fig.~\ref{fig:edgeon}. Because its present-day morphology and kinematics are incompatible with those of the MW, we exclude \texttt{m12r} from further consideration.

\subsection{Building the Merger History} \label{subsec:BuildMerger}

\begin{figure}[t]
    \centering
    \includegraphics[width=1\linewidth]{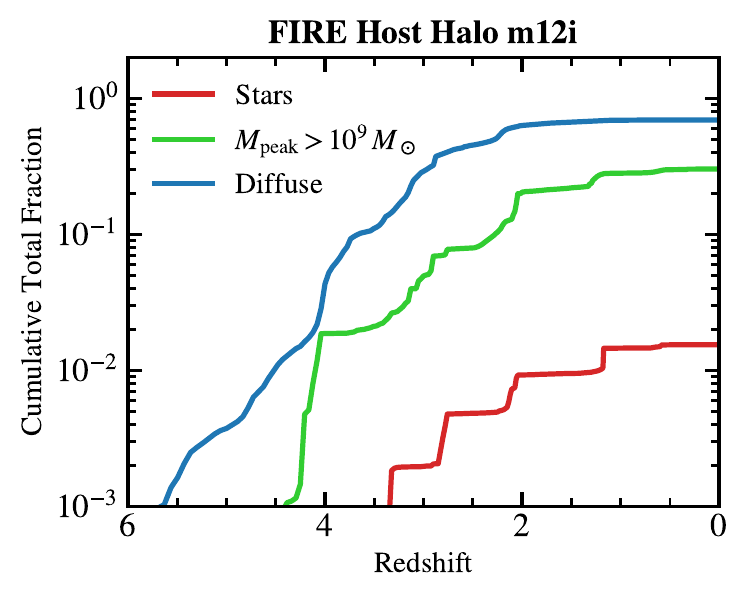}
    \caption{\label{fig:m12i_merger_history} Example accumulated fraction of the stellar and different DM components as a function of redshift for \texttt{m12i}. The red curve represents all accreted stars. The green curve represents DM accreted from major mergers heavier than $10^9~M_{\odot}$ and the blue curve shows the DM accreted either smoothly into the main halo or with unresolved small subhalos ($<10^9M_{\odot}$)}
    
\end{figure}

\begin{table*} [t]
\centering
\begin{tabular}{lcccccccc}
\hline\hline
 & \multicolumn{4}{c}{\textbf{\texttt{m12i}}} & \multicolumn{4}{c}{\textbf{\texttt{m12f}}} \\
\cline{2-5}\cline{6-9}
 & \textbf{I} & \textbf{II} & \textbf{III} & \textbf{IV} & \textbf{I} & \textbf{II} & \textbf{III} & \textbf{IV} \\
\hline
$M_{\rm peak}$ [$M_\odot$] & $6.45\times10^{10}$ & $3.55\times10^{10}$ & $9.64\times10^{9}$ & $3.94\times10^{10}$ & $1.53\times10^{11}$ & $8.20\times10^{10}$ & $3.25\times10^{10}$ & $2.31\times10^{10}$ \\
$\langle{\rm [Fe/H]}\rangle$ (solar circle) & $-1.49$ & $-1.78$ & $-1.84$ & $-2.03$ & $-0.91$ & $-1.16$ & $-1.82$ & $-1.85$ \\
$M_{\rm peak}/M_{\star,\rm total}$ & $68.4$ & $93.4$ & $29.3$ & $277.7$ & $48.0$ & $33.6$ & $161.1$ & $139.9$ \\
$M_{\star,\rm total}$ [$M_\odot$] & $9.43\times10^{8}$ & $3.8\times10^{8}$ & $3.29\times10^{8}$ & $1.42\times10^{8}$ & $3.19\times10^{9}$ & $2.44\times10^{9}$ & $2.02\times10^{8}$ & $1.65\times10^{8}$ \\
$f_\star$ (solar circle) & $0.321$ & $0.281$ & $0.178$ & $0.093$ & $0.441$ & $0.185$ & $0.047$ & $0.038$ \\
$f_{\rm DM}$ (solar circle) & $0.202$ & $0.355$ & $0.020$ & $0.144$ & $0.215$ & $0.357$ & $0.006$ & $0.028$ \\
$z_{\rm acc}$ (mode) & $1.17$ & $2.06$ & $2.76$ & $3.32$ & $0.12$ & $0.73$ & $3.71$ & $4.73$ \\
\hline
\end{tabular}
\caption{\label{tab:m12_mergers}Properties of the most significant accreted mergers in \texttt{m12i} and \texttt{m12f}.
For each merger, we list the peak subhalo mass, the mean stellar
metallicity of accreted stars currently in the solar circle, the peak halo-to-stellar mass ratio, the total stellar
mass of the satellite, the stellar and DM mass fractions contributed within the solar circle (with respect
to the total accreted material from subhalos with $M_{\rm peak} > 10^{9}\,M_\odot$), and the most common accretion
redshift of stripped stars.}
\end{table*}

To construct the merger history, we categorize DM particles into two broad components: major mergers, defined as systems with total mass exceeding $10^9 M_{\odot}$ that contain at least $\sim \mathcal{O}(10)$ star particles, and the diffuse component, which encompasses both smoothly accreted DM and contributions from low-mass subhaloes below the resolution limit. In contrast, star particles either form in situ or are brought in via mergers; since \textit{in-situ} stars are not the focus of this study, we exclusively track those originating from mergers. In this section, we describe our methodology for tracking DM and stellar particles and assigning them to their corresponding progenitors, thereby reconstructing the merger history.

To track the subhalos and identify the particles associated with them, we utilize two complementary packages for halo identification and merger-tree construction: Rockstar and Consistent Trees (CT).
Rockstar is a phase-space halo finder that identifies (sub)halos by clustering particles in six-dimensional phase space and assigning properties such as mass and velocity~\citep{Behroozi_2012a, 2021MNRAS.504.1379S}. Rockstar is run only on the DM particles, and stars are assigned to each halo in postprocessing using similar phase-space criteria. We adopt Rockstar to construct the merger history of each system, as it enables the association of stars with their progenitor subhalos. In each snapshot, Rockstar defines the main halo as the most massive one and thus the galactocentric frame, which allows for a reliable merger catalog out to redshifts of $z \lesssim 3$.

However, defining the galactocentric frame using Rockstar can introduce discontinuities at earlier times. Since the identity of the most massive halo may switch abruptly from one snapshot to the next, particle positions relative to the main halo can exhibit unphysical jumps. These discontinuities may also lead to the misclassification of bound particles as part of the diffuse component.

To mitigate these issues, we instead rely on CT, which builds merger trees by tracing the main progenitor branch of each halo backward in time while enforcing temporal consistency~\citep{Behroozi_2012b}. This results in a smoother and more physically meaningful definition of the main halo center across snapshots. We use CT for two key purposes: (i) computing particle positions relative to the main halo center, and (ii) identifying the accretion of diffuse DM. Specifically, a DM particle is considered accreted once it crosses within the virial radius of the CT-defined main halo.

With the halo catalogs and merger trees in hand, we proceed to reconstruct the merger histories by tracking individual DM and star particles back in time. We begin by selecting all particles located within a solar neighborhood analog at $z=0$, defined as $|r - r_{\odot}| <$ 2~kpc and $|z| \leq$ 1.5 kpc, where r is the spherical radius and $r_{\odot}$ is defined as 8 kpc, as in ~\cite{Necib_2019}. \footnote{Ideally, this cut should be performed in cylindrical coordinates, but the discrepancy is negligible given the focus on the solar neighborhood, which includes a cut to only investigate particles close to the disk.}

For each selected particle, we trace its trajectory through earlier snapshots and determine its subhalo membership. A particle is considered bound to a subhalo at a given snapshot if it lies within the subhalo’s virial radius $R_{200 m}$ and its velocity falls within 2.5$\sigma$ of the subhalo’s internal velocity dispersion. This criterion is applied consistently to both DM and stellar particles. To ensure robust identification, we restrict our analysis to subhalos that contain at least 10 star particles.

Moving backward in time, we identify a DM particle as accreted if it remains associated with the same subhalo in at least six out of the last nine snapshots, including the current one. For star particles, we use a slightly relaxed criterion, requiring membership in three out of the last four snapshots. The difference is due to the star particles needing to be ``born" out of the gas, and therefore cannot be tracked arbitrarily far. This temporal coherence condition minimizes contamination from transient flybys and in-situ particles that temporarily meet the spatial and kinematic criteria. We define the particle’s accretion redshift, $z_{\rm{acc}}$, as the last snapshot at which it is bound to its subhalo\footnote{For consistency checks on the selection criterion please refer to Appendix. \ref{sec:limitation}}. 

For both DM and stellar particles originating from mergers, we record the peak mass of the associated subhalo, the redshift at which the particle was accreted ($z_{\rm{acc}}$), and the particle’s present-day kinematics at snapshot 600. Table \ref{tab:m12_mergers} lists the top four mergers that contributed the greatest fraction of accreted stellar particles at the solar circle for \texttt{m12i} and \texttt{m12f}. An example of the resulting merger history is illustrated for \texttt{m12i} in Figure~\ref{fig:m12i_merger_history}. The red curve represents all accreted stars. The green curve represents DM accreted from major mergers heavier than $10^9~M_{\odot}$ and the blue curve shows the DM accreted either smoothly into the main halo or with unresolved small subhalos ($<10^9M_{\odot}$). Sudden increases in the stellar and major-merger DM fractions correspond to discrete merger events. In contrast, the diffuse component evolves more gradually, with only mild increases near merger times. These modest jumps typically reflect rapid changes in the virial radius following the infall of a massive subhalo, rather than direct particle accretion from the merger itself.

This method is subject to several limitations. The use of both Rockstar and CT can occasionally introduce inconsistencies in the identification of the main halo. At early redshifts ($z \gtrsim 3$), there are instances in which Rockstar and CT disagree on which halo should be designated as the main progenitor. In these cases, two halos of comparable mass exist, and the choice of main branch in the merger tree becomes ambiguous ~\citep[see e.g.,][]{Santistevan_2020}. Because mergers are identified using Rockstar, while diffuse particles are tracked using CT, such disagreements can lead to discrepancies in the reconstruction of early merger events. In rare cases, a halo identified as the main halo by Rockstar may instead appear as merging into the CT main branch. As a result, the classification of mergers occurring before $z \sim 3$ may therefore contain ambiguity.

There are also cases in which our algorithm misidentifies DM particles as belonging to a merger when they do not originate from the accreted subhalo, which we refer to as contamination. For each merger, we estimate this contamination by identifying particles that are classified as belonging to an accreted subhalo, but that were already located within the host halo radius prior to the subhalo’s formal accretion. We then compute the contamination fraction as the ratio of these particles to the total number of particles assigned to the merger. For more massive systems ($M_{peak} \approx 10^{11}M_{\odot}$), contamination from flybys or in situ particles within the main halo can account for up to $10\%$ of particles identified as accreted. For lower-mass mergers ($M_{peak} \approx 10^{10}M_{\odot}$), this contamination fraction is typically reduced to $1–5\%$.  

Additionally, a single physical subhalo is occasionally fragmented into multiple subhalos in the merger tree. This can occur either due to subhalo–subhalo interactions that are not fully resolved into a single object, or as a result of tidal disruption near the main halo, where CT may lose track of the original subhalo and instead assign a new identity. In such cases, we group all associated particles into a single “subhalo group” to ensure consistent treatment in subsequent analyses. More detailed discussion of limitations will be presented in the Appendix.~\ref{sec:limitation}. 

\section{Dark Matter Accreted From Mergers} \label{sec:Correlation}

\begin{figure}[t]
    \centering
    \includegraphics[width=1\linewidth]{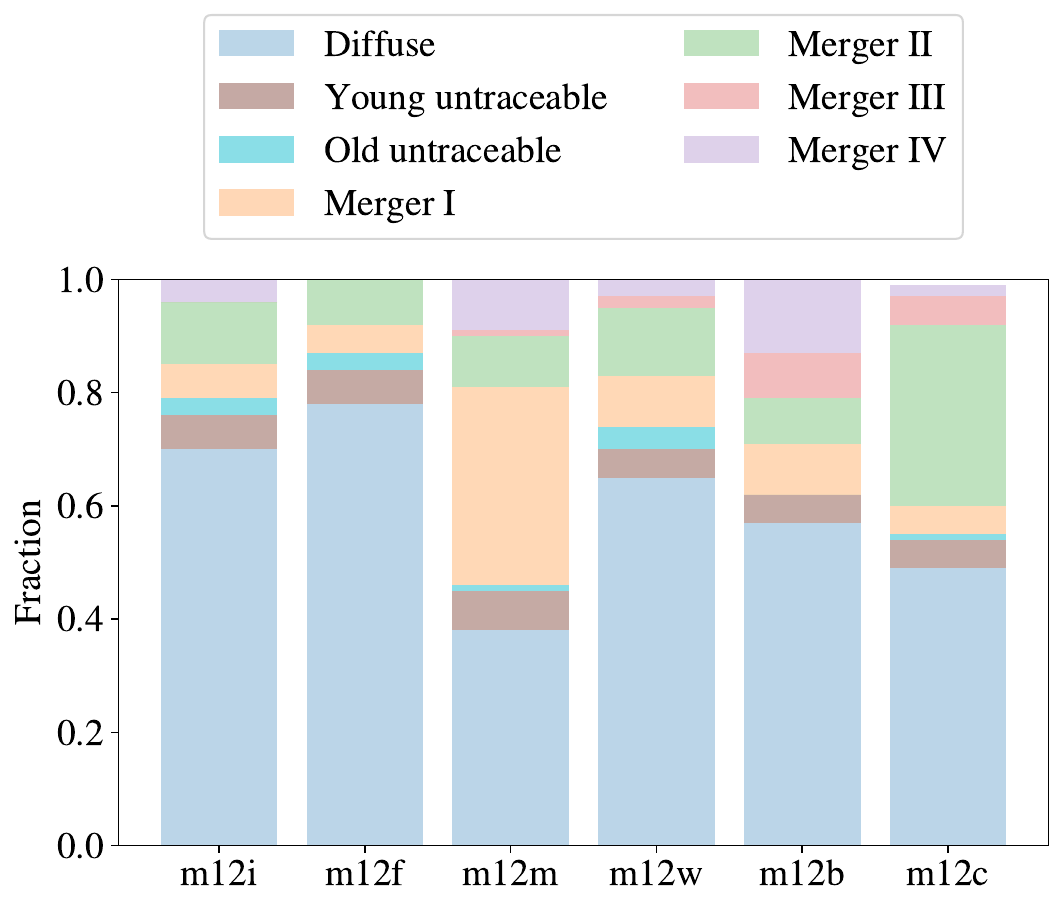}
    \caption{\label{fig:fraction}Fractional contributions of the individual components to the local DM population. The diffuse component is generally the dominant contributor across the \texttt{m12} sample. By contrast, the DM contributions from the top mergers—ranked by their associated stellar mass in the solar neighborhood—vary substantially, and in some cases can be negligible relative to the other components. The exact numerical values of all fractions are listed in Table.~\ref{tab:mergerfractionnew}.
    }
\end{figure}

Mergers enter the host halo as discrete bounded objects at various redshifts. Among these, the most significant contributors to the local stellar population—those that deposit a noticeable fraction of stars into the solar neighborhood—can be tracked using their stellar counterparts. Based on this, we further classify the resolved mergers into three subcategories: (1) the top four mergers that contribute the most stellar mass locally and can be reliably tracked, which we call traceable mergers and are ranked by their contribution to the stellar content in the solar neighborhood, (2) young untraceable mergers that are recent and massive but lack clear stellar signatures, and (3) old untraceable mergers that are too ancient or phase-mixed to be associated with identifiable stellar debris. An illustration of the fractional contributions of the individual components to the local DM population is shown in Fig.~\ref{fig:fraction}.\footnote{The exact values of these fractions are listed in Table~\ref{tab:mergerfractionnew}}.

In this section, we describe how mergers are treated throughout the analysis. We begin by studying traceable mergers in Sec.~\ref{sec:Traceable}. In Sec.~\ref{subsec:EMD}, we introduce the statistical metric used in this work, and in Secs.~\ref{subsec:AH} and \ref{subsec:3d} we apply it to traceable mergers to quantify stellar–DM correlations. We then extend this analysis in Sec.~\ref{subsec:Matching}, where we develop an improved mapping from stellar kinematics to the corresponding DM distributions for each progenitor. Finally, in Sec.~\ref{subsec:Untractable mergers}, we discuss the treatment of untraceable mergers and justify the assumptions adopted for their contributions.

\begin{figure*}[t]
    \centering
    \includegraphics[width=1\linewidth]{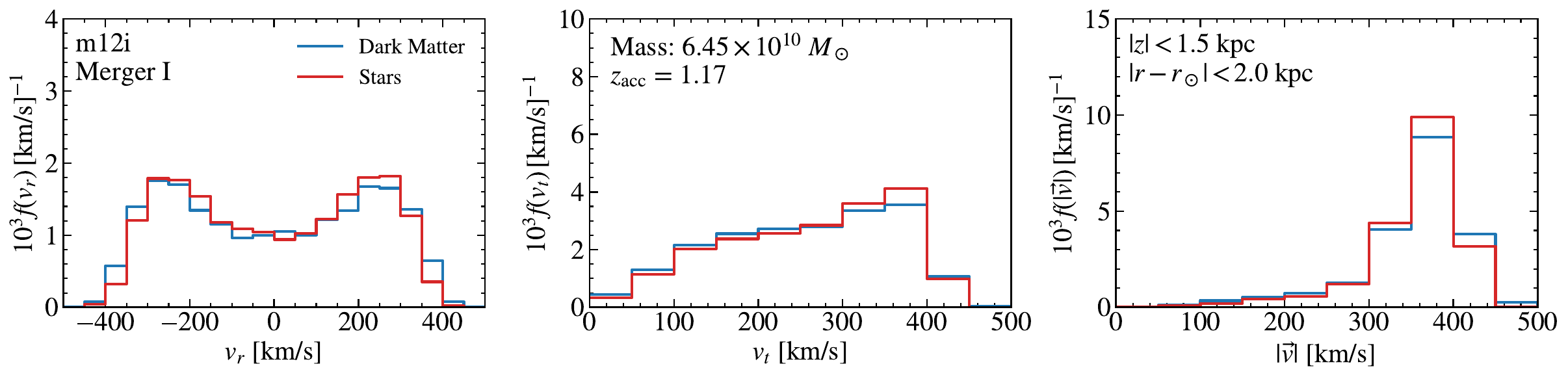}
    \caption{\label{fig:m12i_m1} Example velocity distribution for the largest \texttt{m12i} major merger in radial component $v_r$, transverse component $v_t$ and speed $|\vec{v}|$. It can be noticed that the stellar velocity distributions roughly follow the DM counterpart.}
\end{figure*}

\subsection{The Traceable Mergers} \label{sec:Traceable}
In this section, we investigate the correlation between DM and stellar velocity distributions at redshift $z=0$ for traceable mergers. 
An illustrative example for a traceable merger in \texttt{m12i} is shown in Fig.~\ref{fig:m12i_m1}. The velocities are decomposed into radial and transverse components in spherical coordinates, centered on the galactic center, as well as the total speed distribution for both the stellar and DM components. In this case, the stellar velocity distribution appears to closely follow that of the DM. However, to quantitatively assess the level of agreement, we require a more rigorous statistical measure. To this end, we employ the Earth Mover’s Distance (EMD).

\subsubsection{Earth Mover's Distance} \label{subsec:EMD}

The Earth Mover's Distance, also known as the Wasserstein-1 distance, quantifies how different two distributions are by the minimal work needed to move the probability mass from one to the other, where $work = mass \times distance$~\citep{Villani2003Topics, Villani2009OT}. Formally, for probability measures $P$, $Q$ on $\mathbb{R}^d$~and ground cost/metric $c(x, y)=||x-y||$, 

\begin{equation}
W_1(P,Q)=\inf_{\gamma \in \Pi(P,Q)} \int_{\mathbb{R}^d\times \mathbb{R}^d}c(x, y)d\gamma(x,y), \label{eq:W1}
\end{equation}
where we sum over $\gamma \in \Pi(P,Q)$, with $\Pi(P,Q)$ being the set of joint distributions on the product space whose marginals are $P$ and $Q$ respectively. In 1D, there are convenient closed forms: if $F$ and $G$ are the Cumulative Density Functions (CDFs) of $P$ and $Q$, 
\begin{align} 
W_1(P,Q) &=\int_{-\infty}^{\infty}|F(t)-G(t)|dt \nonumber \\ 
&=\int_0^1|F^{-1}(u)-G^{-1}(u)|du. \label{eq:W1-1d}
\end{align}
Since CDFs are monotonic from 0 to 1, we can integrate the inverse of the function instead, as in the second equation. For empirical samples $\{x_i\}^n_{i=1}, \{y_j\}^m_{j=1}$ sampled from probability distributions $P$ and $Q$ respectively, with optional non-negative weights that sum to 1, $W_1$ is computed by optimally matching the mass along the real line (in 1-D, this reduces to sorting; in higher-D one solves an optimal-transport problem). In our case, we choose uniform weight across all of our samples. 

Because EMD shares the units of the underlying variable (km/s in our application for the EMDs of the velocity distributions), we report a dimensionless effect size 
\begin{equation}
d=\frac{W_1(P,Q)\Phi^{-1}(0.75)}{{\rm{MAD}}(||z||)}, \label{eq:W1-unitless}
\end{equation}
where $\Phi^{-1}(0.75) \simeq 0.67449$ is the 75th percentile of the standard normal distribution,  $\rm{MAD}=median(|X-median(X)|)$ is the Mean Absolute Deviation, and $z\in \{x_i\}^n_{i=1} \cup \{y_j\}^m_{j=1}$. For a normal distribution, this reduces $d$ (from Eq.~\ref{eq:W1-unitless}) to the EMD normalized by the standard deviation $\sigma$. The MAD is preferred over $\sigma$ because it is robust to outliers and heavy tails, while retaining an interpretation as a characteristic half-width. Values of $d \in [0,0.05]$ indicate a small difference between two distributions, while $d\in [0.05, 0.2]$ indicates a small-to-moderate difference. In practice, we normalize by the MAD of the stellar velocity distribution rather than the pooled stellar-plus-DM distribution, so that $d$ can be read as the number of observed standard deviations separating the DM and stellar distributions.\footnote{We have verified that using the pooled MAD yields only marginally smaller values of $d$.}

\subsubsection{Accretion History} \label{subsec:AH}

\begin{figure*}[t]
    \centering
    \includegraphics[width=1\linewidth]{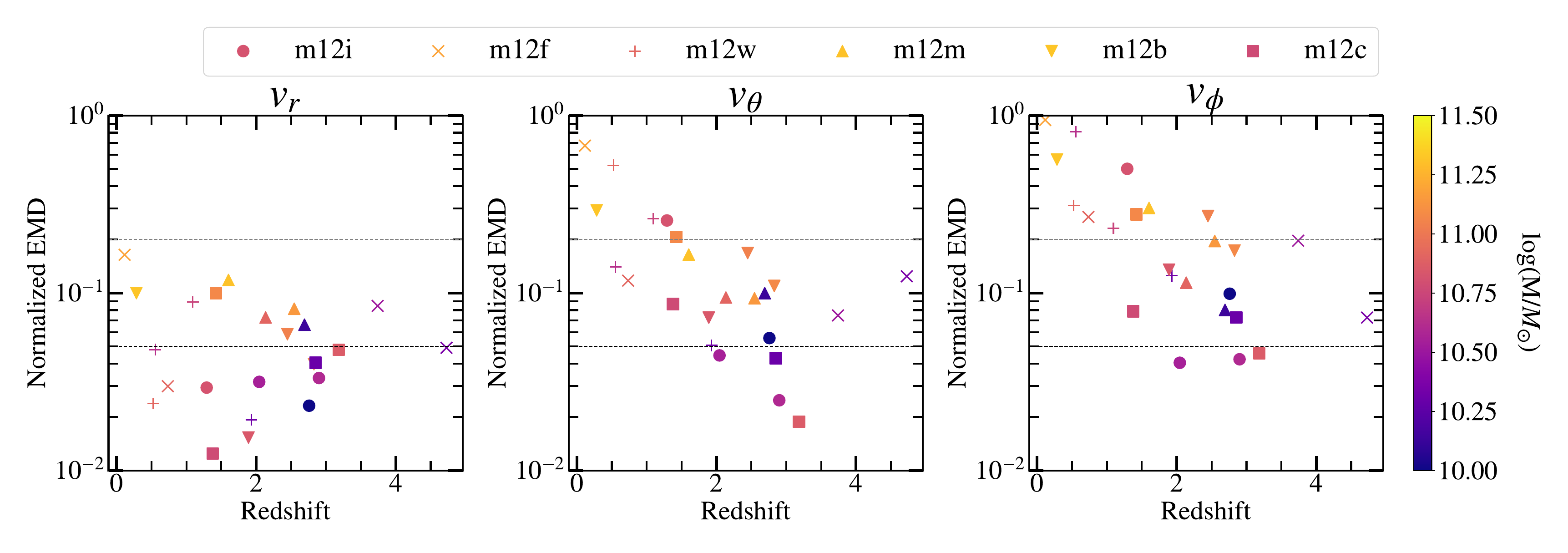}
    \caption{\label{fig:EMD-RTS} EMD for $v_r$, $v_{\theta}$, and $v_{\phi}$ between the stellar and DM velocity distributions with respect to the host disk for the top four mergers of all six of our \texttt{m12} galaxies as a function of redshift. Each marker shape stands for a different \texttt{m12} and is colored by the merger mass. The horizontal dashed lines at EMD=0.2 and EMD=0.05 mark the thresholds for meaningfully different and nearly identical distributions, respectively. It can be seen that the stellar and DM radial velocities are similar to each other due to violent relaxation. The difference in speed is thus mainly driven by the transverse component. The transverse EMD decreases with redshift, reflecting the phase mixing of earlier mergers.
    }
\end{figure*}

Equipped with the Earth Mover’s Distance (EMD) framework, we now apply this statistic to our full merger sample. Specifically, we focus on the four mergers that contribute the largest stellar mass to the solar neighborhood in each of the six \texttt{m12} simulations, resulting in a total of 24 mergers. The corresponding results are shown in Fig.~\ref{fig:EMD-RTS} as scatter plots. For each merger, we compute the EMD between the stellar and DM distributions for each velocity component within the solar neighborhood. The horizontal axis shows the accretion redshift $z$, with each simulation denoted by a distinct marker, color-coded by the merger's peak mass. The vertical axis shows the normalized EMD, interpreted as the difference between the DM and stellar velocity distributions in units of the characteristic statistical uncertainty (i.e., an effective number of standard deviations). The horizontal dashed lines at EMD$=0.2$ and EMD$=0.05$ indicate the thresholds for meaningfully different and nearly identical distributions, respectively

As shown in Fig.~\ref{fig:EMD-RTS}, the radial velocity components of the stellar and DM populations are generally similar regardless of accretion time or progenitor mass. All components exhibit a dependence on these quantities, but the $v_\theta$ and $v_\phi$ components show a stronger correlation: DM and stellar distributions are more closely matched for earlier, lower-mass mergers than for more recent, massive ones. We note, however, that due to the hierarchical growth of cold DM halos, earlier mergers also tend to be less massive, as reflected in the colors of Fig.~\ref{fig:EMD-RTS}.

This behavior is physically well motivated. As satellites are accreted, they undergo violent relaxation and phase mixing predominantly along the radial direction, where the host potential varies most rapidly~\citep{1967MNRAS.136..101L, 1986MNRAS.219..285T, Tremaine_1999}. The transverse velocity, by contrast, is tied to the orbital angular momentum, which is approximately conserved and therefore relaxes on a longer timescale. Differences in the total speed distributions are consequently driven primarily by discrepancies in the transverse components.

Moreover, because stars are concentrated at the center of their progenitor's potential well and do not extend as far as the DM, the outer DM is stripped first while the inner stellar component is disrupted later. The stellar velocity distribution therefore lags behind the DM in terms of phase mixing: the radial components of both populations equilibrate relatively quickly, while the stellar transverse component relaxes more slowly. Given sufficient time after infall, the accreted material becomes fully virialized and the stellar–DM correlation strengthens.

Consistent with this picture, we find smaller EMD values for lower-mass mergers compared to higher-mass ones accreted at the same time, because lower-mass mergers are more easily disrupted, whereas more massive subhalos retain extended DM envelopes that are stripped well before their inner stellar cores.

\subsubsection{Robust Comparison on Full Velocity Distributions\label{subsec:3d}}

\begin{figure}[t]
    \centering
    \includegraphics[width=1\linewidth]{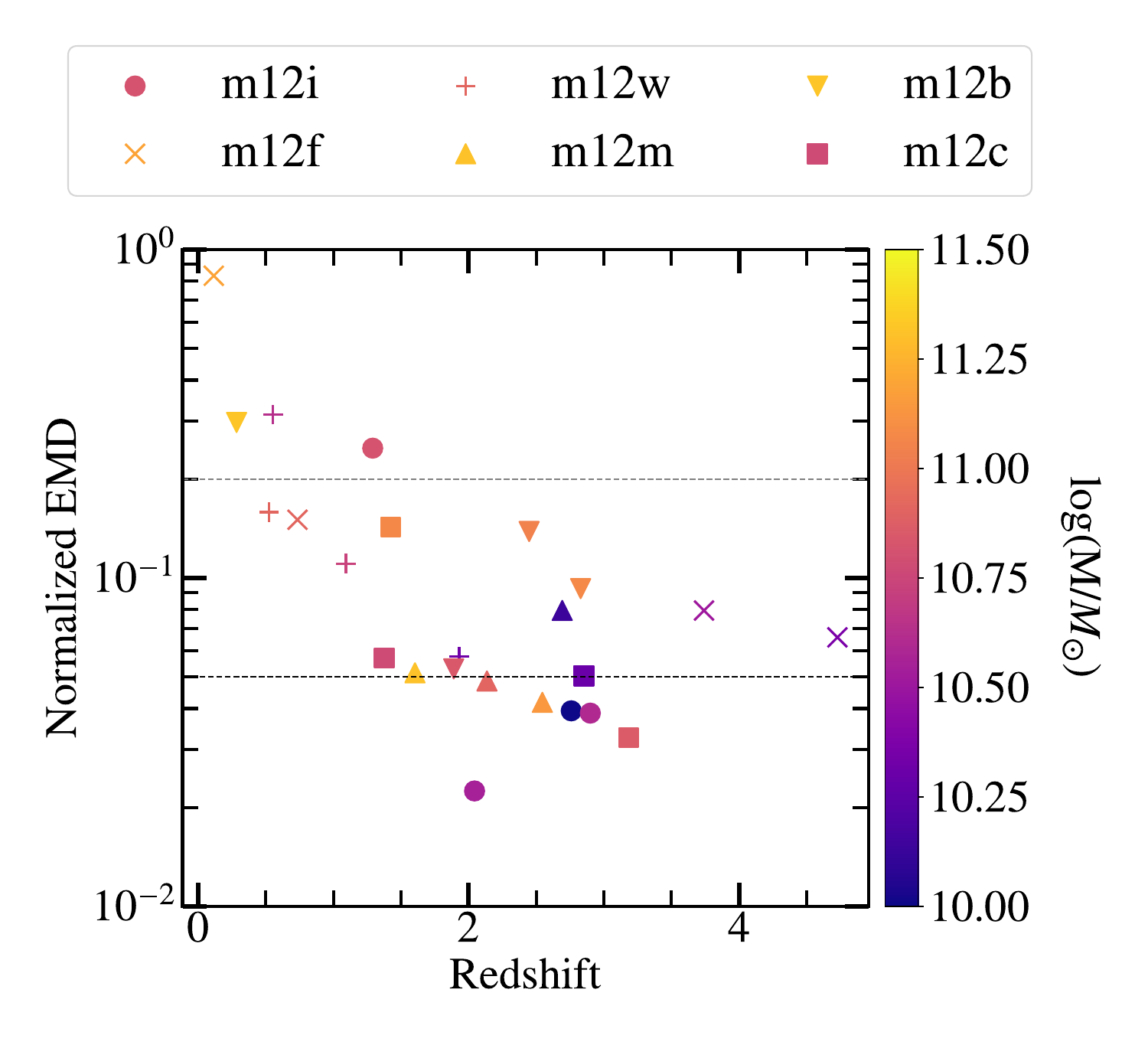}
    \caption{\label{fig:EMD-3d}Coordinate-independent EMD for the full 3d velocity distribution. Again, a clear trend can be seen that decreases with redshift, which is consistent and mainly driven by the angular velocity. The standard error is estimated by the standard deviation of the random draws divided by square root of K. The errorbars are all quite small and thus not included in the plot.}
\end{figure}

For conventional, non-directional direct-detection experiments with elastic, velocity-independent scattering, the differential event rate depends only on the DM speed distribution in the lab frame, $f_{\rm{lab}}(v)$, rather than on the full three-dimensional velocity vector. One might therefore imagine comparing only the one-dimensional speed distributions of stars and DM. However, the lab-frame speed distribution is obtained by boosting the Galactocentric three-dimensional velocity distribution by the Solar and terrestrial motions~\citep[see e.g.][]{Jungman_1996}. This mapping is sensitive to the anisotropy and substructure of the underlying halo; distinct three-dimensional distributions can yield similar speed histograms in the Galactic frame but lead to different speed distributions and modulation signals in the lab frame. Moreover, in more general non-relativistic effective field theory treatments of DM–nucleus scattering, many operators carry explicit dependence on the velocity (e.g. through $v_{\perp}$) and on the momentum-transfer direction, so the full three-dimensional velocity distribution, not just its speed marginalization, is physically relevant \citep{Fitzpatrick_2013, anand2014modelindependentanalysesdarkmatterparticle, Kavanagh_2015}. 

So far we have only been working on the velocity components separately. It is difficult to combine the individual EMDs and get a meaningful and self-contained result, since the component-wise EMDs are coordinate-dependent. One way of course is to follow the formal definition in eq.~\ref{eq:W1} and compute the full cost matrix $c(x, y)$ for each pair of data points and solve the optimal-transport problem. However, it will be a higher dimensional optimal-transport problem with the cost matrix scaling as the product of the numbers of star $n$ and DM particles $m$ involved $\mathcal{O}(n\times m)$. So the computational memory cost grows quickly with the number of particles. We thus resort to a computationally more efficient test, the sliced Wasserstein~\citep{Bonneel2015SlicedRadonBarycenters, Nietert2022SlicedWassersteinGuarantees}. 

Sliced Wasserstein (SW) sidesteps the computational cost by picking a random direction on a unit sphere, $\theta \in S^2$, projecting each 3d data point onto that direction, computing 1d Wasserstein between the two projected 1d samples and taking the expectation value with respect to that direction. Formally, let $X=\{x_i\}^n_{i=1}, Y=\{y_j\}^m_{j=1} \subset \mathbb{R}^3$, and $\theta \in S^2$, and the projections $X_{\theta}=\{\langle \theta, x_i\rangle\}, Y_{\theta}=\{\langle \theta, y_j\rangle\}$. Then SW is 
\begin{equation}
SW_1(X,Y)= \mathbb{E}_{\theta \in  S^2} [W_1(X_{\theta}, Y_{\theta})].\label{eq:SW1}
\end{equation}
Thus, SW is just sorting and integrating CDF differences, making the computational scaling  $O(m+n)$. The expectation value scales as the sum of the integrals over all directions and thus is invariant under rotation and captures the information of the full 3d structure. In practice, the expectation is approximated by a Monte Carlo average over $K$ random directions. The systematic error comes from Monte Carlo random draws and scales as $1/\sqrt{K}$. Thus we choose $K=512$ in our case so that the systematic error is less than $5\%$. We then normalize SW as before with $\Phi^{-1}(0.75)/\rm{MAD}(||v||)$ to get our robust, dimensionless, coordinate-independent measure of similarity between two full 3-d velocity distributions. 

The resulting plot is shown in Fig.~\ref{fig:EMD-3d}. Even if the transverse velocity distributions are in general less correlated than radial components, the whole 3d velocity distributions of DM and the stars are similar for subhalos merging before $z\sim1.5$. We also notice that although the latest merger in \texttt{m12i} has small normalized EMD values for all of its components separately, its SW is much higher. This is due to the fact that even though its transverse component magnitude roughly follows the same distribution, the direction of the transverse velocity vectors is different. It is also a prime example of how component-wise EMD alone is not a good characterization of similarity between two speed distributions. 

\subsubsection{Matching stellar velocity distribution with DM}\label{subsec:Matching}

\begin{figure*}[t]
    \centering
    \includegraphics[width=1\linewidth]{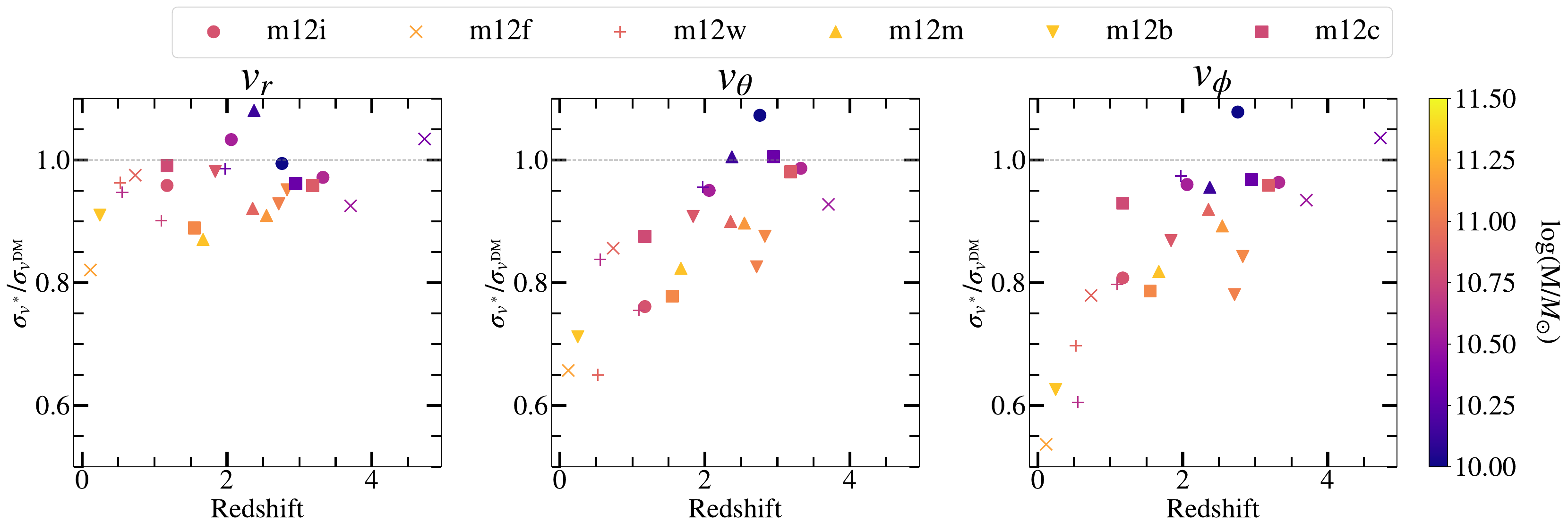}
    \caption{\label{fig:dispersion}Ratio of the stellar velocity dispersion to the dark-matter velocity dispersion for particles accreted from the same progenitor, shown as a function of accretion redshift. Recent mergers tend to exhibit broader dark-matter velocity distributions than their stellar counterparts (ratio $<1$), while the discrepancy decreases for earlier accretion times, consistent with increased virialization and phase mixing in older merger debris. }
\end{figure*}

\begin{figure}[t]
    \centering
    \includegraphics[width=1\linewidth]{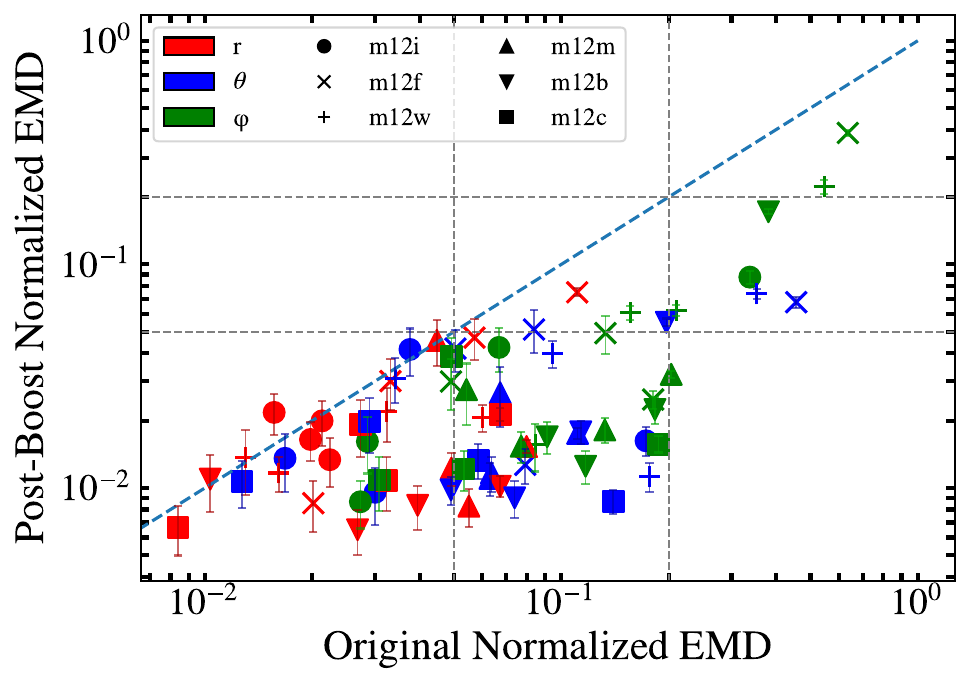}
    \caption{\label{fig:EMD-Improved}Comparison of EMD before and after applying Gaussian convolution to the stellar velocity distributions. Each point corresponds to a merger and velocity component, with the $x$-axis showing the pre-convolution EMD and the $y$-axis showing the post-convolution EMD. The gray dashed lines indicate reference values at EMD~$=0.05$ and EMD~$=0.2$. All points lie on or below the diagonal line, confirming that the convolution procedure either improves or preserves the original fit. Nearly all post-convolution EMD values fall below the EMD~$=0.2$ threshold, indicating significantly improved agreement between the stellar and DM velocity distributions after convolution, particularly for the angular components.}
\end{figure}

\begin{figure*}[t]
    \centering
    \includegraphics[width=1\linewidth]{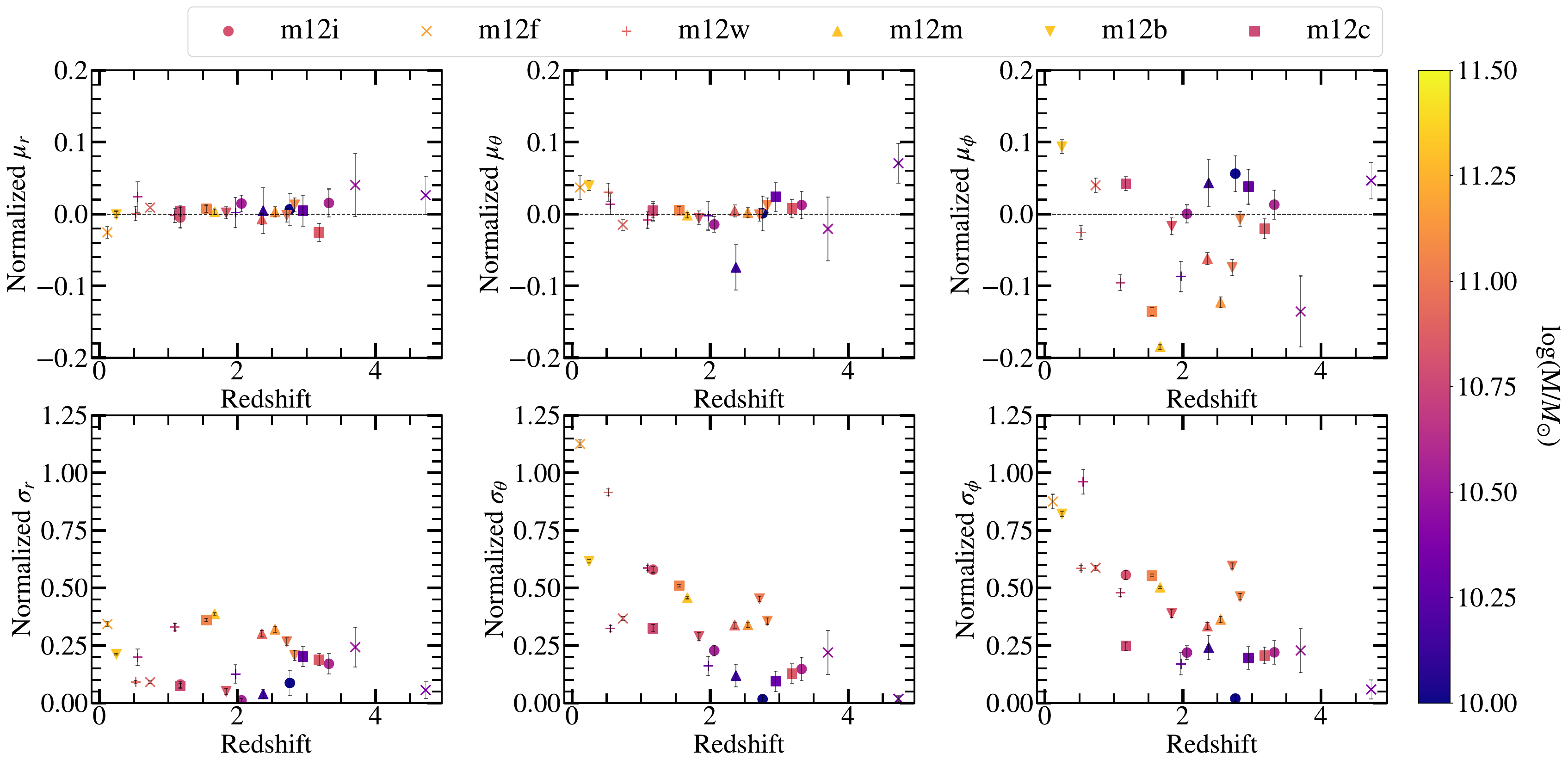}
    \caption{\label{fig:mu} Best-fit values of $\mu$ and $\sigma$, normalized by the stellar median absolute deviation (MAD), plotted as a function of merger redshift. The normalized $\mu$ values cluster around zero across a wide range of redshifts and progenitor masses, indicating that a systematic shift between the stellar and DM velocity distributions is generally unnecessary when modeling the angular components. In contrast to $\mu$, the normalized dispersion exhibits a clear dependence on redshift for the angular components, with larger values at later accretion times. This trend is well described by a decaying exponential function of redshift. }
\end{figure*}

In previous sections, we quantified the correspondence between the local DM and stellar velocity distributions associated with mergers, both component-wise and for the full 3-dimensional distribution. Here, we present a method for approximating the velocity distribution of accreted DM in major mergers using the kinematics of its associated stellar counterpart. We convolve the stellar velocity distribution with a Gaussian function to better match with the DM counterpart accreted from each merger and establish a scaling relation of the best fit parameters across the simulations as a function of redshift so that it can be generalized to observation. 

Although accreted stars and DM are expected to follow similar phase-space distributions due to their shared origin, in practice we observed that the stellar velocity distributions are typically more concentrated than their DM counterparts. Fig.~\ref{fig:dispersion} shows the ratio of the stellar velocity dispersion to that of the DM accreted from the same progenitor. For more recent mergers, the DM distribution is systematically broader than its stellar counterpart, while this discrepancy diminishes toward higher accretion redshift. This trend is consistent with the expectation that earlier mergers have had more time to virialize and phase-mix, leading to more similar stellar and dark-matter kinematics. This arises because stars reside deeper in the subhalo potential wells and are therefore disrupted later than the more extended DM components. As a result, directly using the stellar velocity distribution to approximate that of DM — for instance, by matching speed distributions — tends to underestimate the width of the DM distribution.

Hence, to account for the mismatch, we convolve the stellar velocity distribution with a Gaussian kernel $N(\mu, \sigma)$ component-wise.
\begin{equation}
    f^{\rm{DM}}_i = f_i^**N(\mu, \sigma),
\end{equation}
where $f^{\rm{DM}}_i$ and $f_i^*$ are the local DM and stellar velocity distributions accreted from the $i$-th merger respectively. The convolution broadens the stellar velocity distribution to better match the corresponding DM distribution.\footnote{Technically, the distribution can also shift via $\mu$. However, our results show that $\mu\sim0$.} To determine the best-fit parameters, we compute the EMD between the convolved stellar distribution and the true DM distribution, and select $\mu$ and $\sigma$ to minimize the post-convolution EMD. 

By construction, this procedure guarantees that the EMD can only improve or remain unchanged. In the limit where the optimal kernel has $\mu=0$ and $\sigma=0$, the convolution reduces to the identity operation and recovers the original EMD. Thus, the method ensures that the convolved distribution is at least as good a match to the DM velocity distribution as the original stellar one. Since the radial velocity distributions of accreted stars and DM are already closely aligned, we apply the Gaussian convolution only to the angular components (i.e., the tangential velocity directions), where the discrepancies are more significant. 

The impact of this procedure is shown in Fig.~\ref{fig:EMD-Improved}, where we plot the post-convolution EMD against the pre-convolution EMD for each merger and velocity component. Gray dashed lines indicate the reference values of EMD=0.05 and EMD=0.2.\footnote{Again, one can think of these EMD values as fractions of standard deviations.} All points lie on or below the diagonal, confirming that the convolution either improves or preserves the original fit. Moreover, nearly all post-convolution EMD values fall below the 0.2 threshold, with only a couple of exceptions, indicating that the convolved stellar distributions provide a better approximation to the corresponding DM velocity distributions in most cases.

We plot the best fit values of $\mu$ for each merger as a function of redshift in Fig.~\ref{fig:mu}. To enable meaningful comparisons across mergers, we normalize $\mu$ by the stellar MAD as in previous sections. The majority of the normalized $\mu$ values cluster around zero, with no significant dependence on redshift or progenitor mass. This suggests that a systematic \textit{shift} between the stellar and DM velocity distributions is generally unnecessary.

The best fit dispersion of the boosted Gaussian $\sigma$, also normalized by stellar MAD, is shown in the bottom row of Fig.~\ref{fig:mu}. For the radial component, the normalized best fit $\sigma$ is small ($\lesssim 0.3$) in general which is expected since the difference in the velocity dispersion is small as shown in Fig.~\ref{fig:dispersion}. In contrast, $\sigma$ of the angular components exhibit a clear trend with redshift, increasing at later times. Thus a gaussian convolution in the radial component is unnecessary, and we will only apply the convolutions to the angular components in the following analysis. 

We find that this trend in the angular components is well captured by fitting $\sigma$ as a decaying exponential function of redshift using \texttt{scipy}~\citep{Virtanen2020SciPy}. Specifically, the normalized dispersion can be approximated by the relations
\begin{equation}
\begin{split}
    \sigma_{\theta}&= 0.72~e^{-0.30~z}, \\
    \sigma_{\phi}&= 0.79~e^{-0.30~z}.
\end{split}
\end{equation}

A related approach was independently developed by \cite{folsom2025darkmattervelocitydistributions}, who also broaden the stellar velocity distribution to approximate the DM counterpart. The two methods differ in a few respects. First, \cite{folsom2025darkmattervelocitydistributions} apply an affine transformation to the stellar distribution (their equations 5 and 6), introducing a Gaussian kernel at a later stage as a smoothing step, whereas our framework convolves the stellar distribution directly with a Gaussian kernel. Second, their treatment applies a single velocity-dispersion boost averaged across all components, while ours models each component separately and characterizes the stellar–DM mismatch as a function of merger redshift. Third, their boost factors are reported in physical units of km/s, while ours are normalized by the stellar MAD, making them dimensionless and therefore more readily comparable across mergers of different mass scales.

\subsection{Untraceable mergers}\label{subsec:Untractable mergers}

The untraceable mergers are defined in two categories: first, the young untraceable, as mergers more massive than $10^9 M_{\odot}$ in halo mass but with little contribution to the local stellar mass, and second, the old untraceable, as those accreted earlier than redshift 3 which is more ancient and phase-mixed. 

\begin{figure*}[t]
    \centering
    \includegraphics[width=1\linewidth]{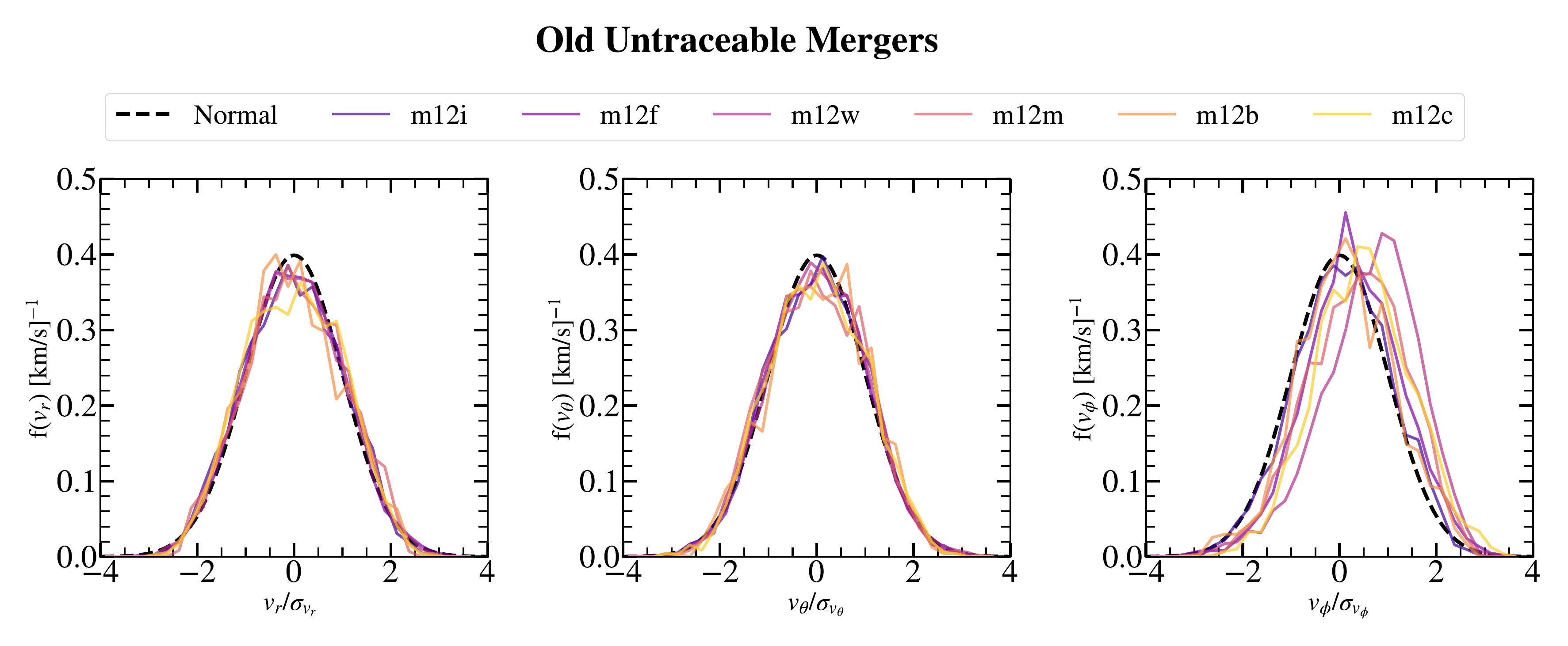}
    \includegraphics[width=1\linewidth]{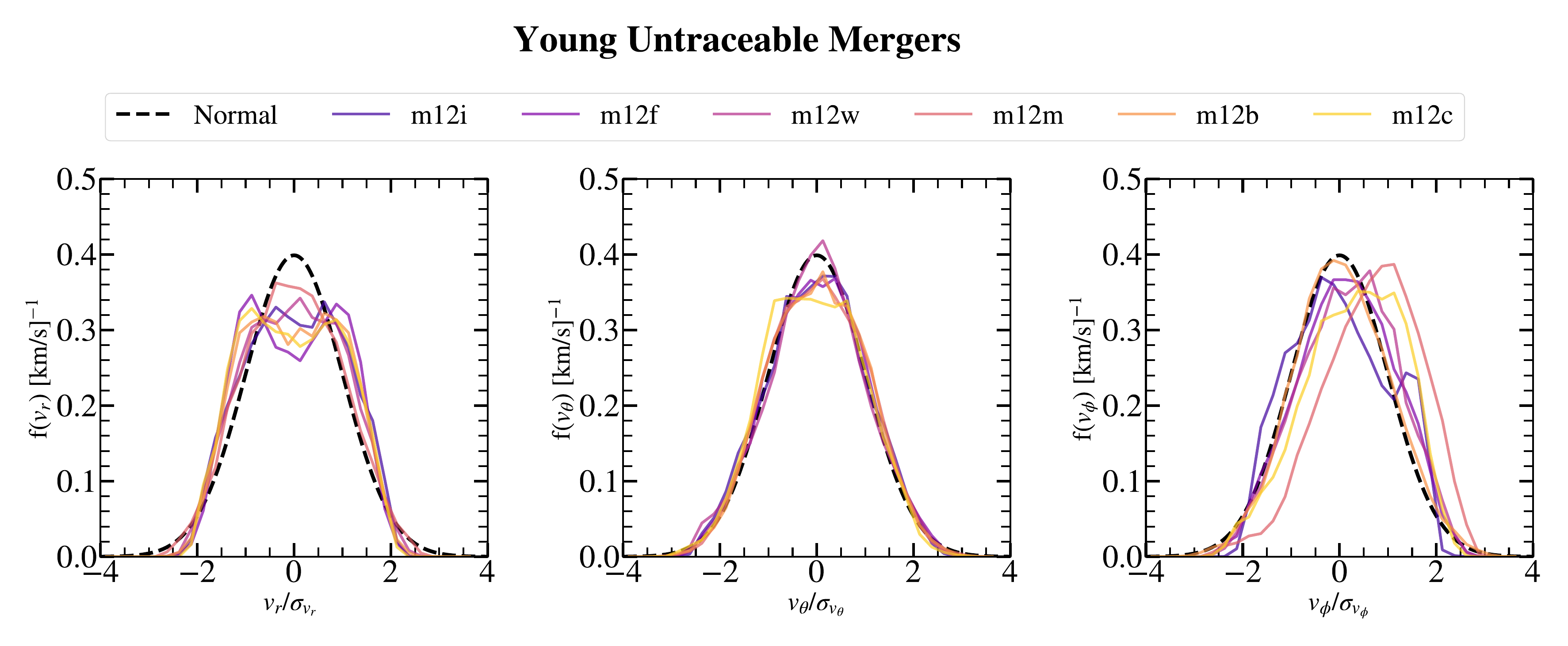}
    \caption{\label{fig:untractables}Velocity distributions of untraceable mergers, separated into old and young populations. Old untraceable mergers exhibit velocity components that are approximately Gaussian, with the exception of \texttt{m12w}, which shows noticeable co-rotation. In contrast, younger untraceable mergers deviate significantly from Gaussian due to incomplete dynamical relaxation. These distributions are noisy and irregular due to limited particle statistics, making precise modeling challenging. As they contribute less than $10\%$ of the total local DM mass (see Table~\ref{tab:mergerfractionnew}) , we approximate them as part of the diffuse component and treat any discrepancies as systematic uncertainties.}
\end{figure*}

Their velocity distributions are shown in Fig.~\ref{fig:untractables}. The old untractable mergers can be well approximated by a Gaussian distribution in all of their components except for \texttt{m12w} which has a non-trivial co-rotation. The young untraceable mergers, by contrast, deviate more strongly from a Gaussian form, consistent with their incomplete dynamical relaxation. Nevertheless, these distributions appear noisy and irregular due to limited particle statistics, making them difficult to model precisely. As shown in Table.~\ref{tab:mergerfractionnew}, they only make up sub $10\%$ of the total mass. So we will assume them to behave the same as the diffuse component which will be discussed in the next section, and treat any other discrepancy as a systematic uncertainty. 

\section{The Diffuse Component}\label{sec:Diffuse}

\begin{table*}[t]
\centering
\begin{tabular}{lcccccccc}
\hline\hline
 & \multicolumn{2}{c}{\textbf{Fiducial}} & \multicolumn{2}{c}{\textbf{Low Resolution}} & \multicolumn{2}{c}{\textbf{High Resolution}}\\
\cline{2-3}\cline{4-5}\cline{6-7}
 & \textbf{$\beta$} & \textbf{$\alpha$/$v_c$} & \textbf{$\beta$} & \textbf{$\alpha$/$v_c$} & \textbf{$\beta$} & \textbf{$\alpha$/$v_c$}\\
\hline
$r$ & 2.74 $\pm$0.31 & 1.17 $\pm$0.11 & 2.51 $\pm$0.14 & 1.30 $\pm$0.06 & 2.55 $\pm$0.03 & 1.15 $\pm$0.01\\
$\theta$ & 2.37 $\pm$0.20 & 1.05 $\pm$0.07 & 2.40 $\pm$0.16 & 1.10 $\pm$0.06 & 2.55 $\pm$0.03 & 1.01 $\pm$0.01 \\
$\phi$ & 2.35 $\pm$0.22 & 1.08 $\pm$0.09 & 2.36 $\pm$0.27 & 1.23 $\pm$0.08 & 2.16 $\pm$0.05 & 1.12 $\pm$0.01 \\
\hline
\end{tabular}
\caption{\label{tab:best_fit}Best-fit generalized Gaussian parameters for the diffuse DM velocity distribution. The location parameter is consistent with $\mu \simeq 0$ for all components, so we report the remaining shape parameters (e.g., $\alpha$ and $\beta$) for each velocity component. The same fitting is also applied to low- and high-resolution simulations. }
\end{table*}

The other major contributor to the local DM population is the diffuse component. This component includes both DM smoothly accreted onto the host halo and unresolved contributions from low-mass subhaloes below the resolution limit of the simulation (i.e. might not have enough star particles to get correctly traced). For the \textit{Latte} suite, we define the resolution threshold for resolved subhalos as $10^9M_{\odot}$ in total mass. This threshold is scaled proportionally with the DM particle mass in the lower- and higher-resolution versions of each simulation, as we discuss in our resolutions tests (Sec.~\ref{subsec:Resolutions}).

\begin{figure}[t]
    \centering
    \includegraphics[width=1\linewidth]{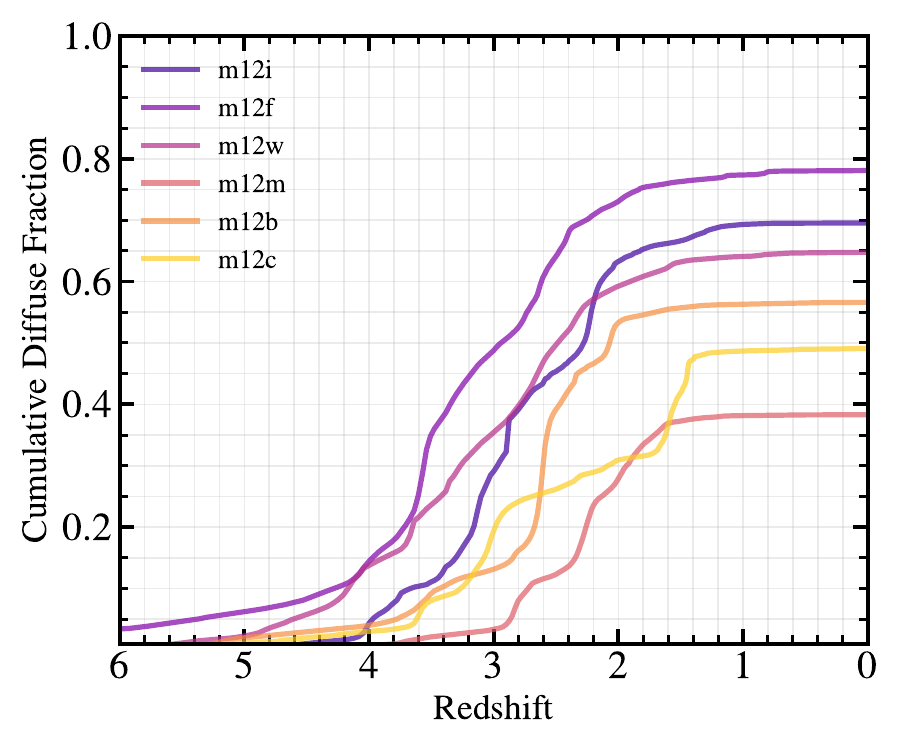}
    \caption{\label{fig:diffuse_fraction} Cumulative growth of the diffuse DM component as a function of accretion redshift, defined by the methods in Sec.~\ref{subsec:BuildMerger}. The diffuse component can contribute between $40\%$ and $80\%$ of the local DM, depending on the specific accretion history. The accretion history is generally smooth, with small jumps coinciding with major merger events (see the text for more discussion).}
\end{figure}

\begin{figure*}[t]
    \centering
    \includegraphics[width=1\linewidth]{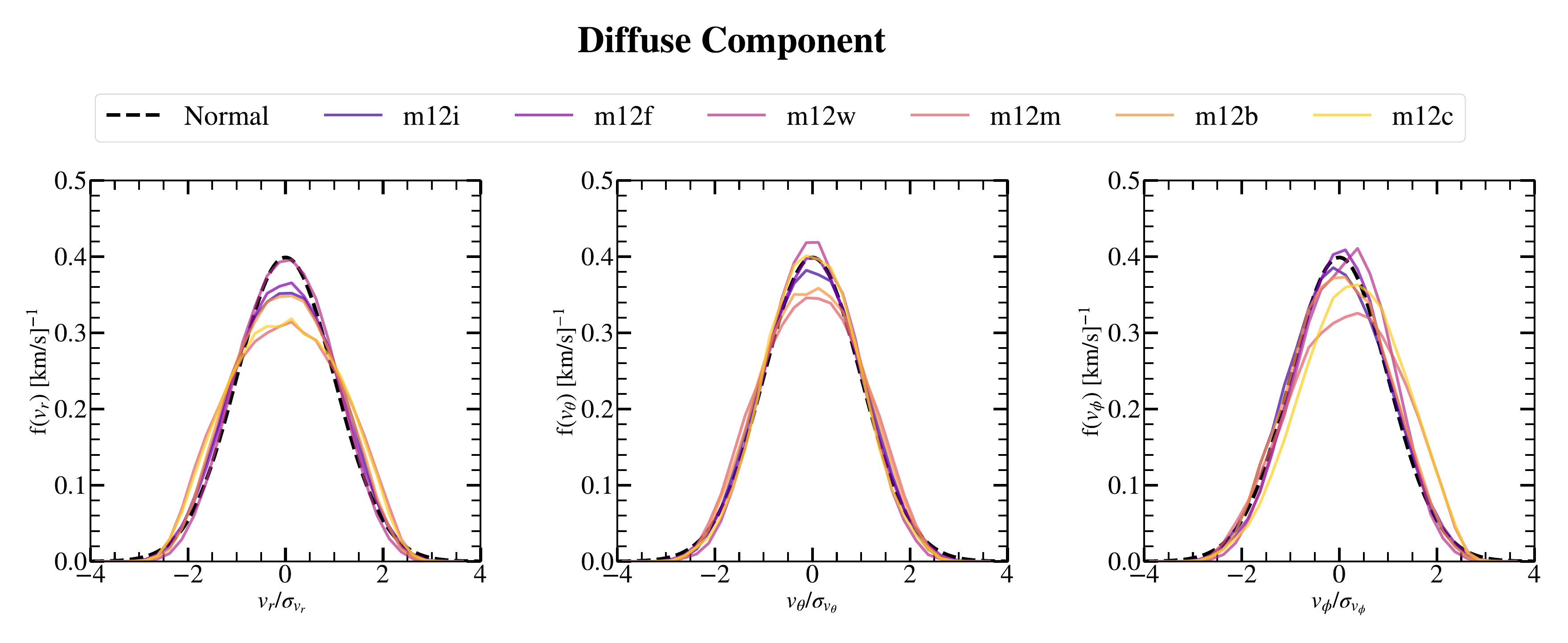}
    \caption{\label{fig:diffuse_velocity}Normalized velocity distributions of the diffuse DM component, shown for each velocity component. While the standard Gaussian approximation (as assumed in the Standard Halo Model) provides a reasonable fit, systematic deviations are evident — particularly in the peak sharpness and the tails. These discrepancies arise due to ongoing perturbations from mergers and deviations from spherical symmetry induced by the galactic disk.}
\end{figure*}

The diffuse component constitutes a substantial fraction of the local DM population—often exceeding the contribution from resolved mergers. The exact fraction varies significantly depending on the specific accretion history of the host halo and can range from approximately $40\%$ to as high as $80\%$ (see Tab.\ref{tab:mergerfractionnew}). As shown in Fig.~\ref{fig:diffuse_fraction}, the diffuse component typically exhibits a smoothly growing profile over time, with modest jumps that coincide with the accretion times of major mergers. These features do not solely imply significant misclassification of merger-associated DM particles as diffuse as we tested in Appendix.~\ref{sec:flyby}, but rather reflect the rapid growth of the host's virial radius during major accretion events, which causes more particles to fall within the host halo boundary and be tagged as diffuse.

The diffuse component of the DM halo is fundamentally untraceable by stellar proxies, as it lacks any associated luminous counterpart. As the dominant and most smoothly accreted component of the halo, it most closely satisfies the assumptions underlying the Standard Halo Model — approximate spherical symmetry and isotropy — which predict a Gaussian velocity distribution. However, continuous perturbations from infalling substructures and the formation of the galactic disk break these symmetries and prevent full thermalization, so there is no guarantee that a Gaussian is adequate, as shown in Fig.~\ref{fig:diffuse_velocity}. 

To allow for departures, we fit the diffuse velocity distribution with a generalized Gaussian,
\begin{equation} \label{eq:generalized_gaussian}
f_g(v) =
\frac{\beta}{2 \alpha \, \Gamma\!\left(1/\beta\right)}
\exp\!\left[
-\left|\frac{v - \mu}{\alpha}\right|^{\beta}
\right],
\end{equation}
where $\mu$ denotes the location parameter, $\alpha > 0$ is the scale parameter, and $\beta>0$ governs the shape of the distribution. The standard normal distribution is recovered when $\beta = 2$, with $\mu$ as the mean and $\alpha/\sqrt{2}$ as the standard deviation; for $\beta > 2$, the distribution becomes more boxy, with a narrower core and suppressed tails. As shown in Fig.~\ref{fig:diffuse_velocity}, the best-fit shape parameters prefer $\beta > 2$, indicating that the diffuse velocity distribution is systematically more concentrated than the Gaussian assumed by the SHM.

\begin{figure}[t]
    \centering
    \includegraphics[width=1\linewidth]{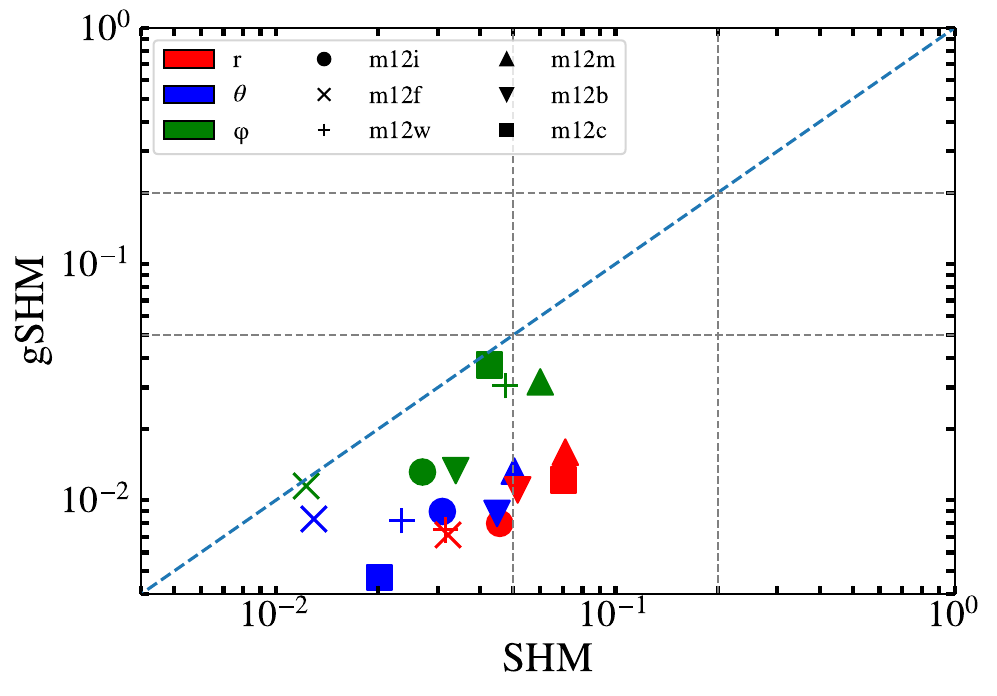}
    \caption{\label{fig:diffuse_EMD}Comparison of EMD between best-fit Gaussian and generalized Gaussian models for the diffuse DM velocity distributions. Each point corresponds to a velocity component from a given simulation, with different marker shapes denoting different simulations and colors representing velocity components. All points fall below the y=x line, indicating that the generalized Gaussian consistently achieves lower EMD values than the Gaussian fit, and thus provides a more accurate representation of the diffuse velocity distribution.}
\end{figure}

We fit the diffuse DM component using both a Gaussian and a generalized Gaussian model, and quantify the quality of each fit by computing the EMD between the best-fit model and the simulated velocity distribution. The resulting EMD values for the Gaussian and generalized Gaussian fits are compared in Fig.~\ref{fig:diffuse_EMD}. Different simulations are indicated by distinct marker shapes, while different velocity components are shown in different colors. All data points lie well below the $y=x$ line, demonstrating that the generalized Gaussian consistently yields a lower EMD than the Gaussian fit. Although this model has one extra degree of freedom, the consistent improvement across simulations and components—often well beyond small incremental changes expected from increased parameterization—indicates that the diffuse velocity distribution is intrinsically non-Gaussian. In particular, we find that the best-fit location parameter is consistent with zero, while the remaining shape parameters are presented in Table.~\ref{tab:best_fit}. 

It is evident that the velocity distributions of the diffuse component consistently deviate from a standard Gaussian, with the radial component showing the largest departure. This is due to the interplay between mergers and the host halo. As mergers are accreted, the associated time-dependent fluctuations in the host gravitational potential drive violent relaxation and redistribute the energies and orbital structure of the pre-existing dark matter~\cite{Sparre_2012, Ogiya_2016, PhysRevD.99.023012}. Such episodes preferentially affect the radial degree of freedom, broadening and reshaping the $v_r$ distribution relative to the equilibrium Gaussian expectation and thereby enhancing the non-Gaussian features observed in the diffuse component. We also find that the best-fit scale parameters $\alpha$ — for both the Gaussian and generalized Gaussian models — are slightly larger than the local circular velocity $v_c$, indicating broader velocity distributions than expected under the SHM. In the Gaussian case, this implies that the inferred velocity dispersion $\sigma = \alpha/\sqrt{2}$ exceeds the SHM-predicted value of $v_c/\sqrt{2}$, further reinforcing that the diffuse component exhibits modestly enhanced kinematic dispersion compared to the SHM baseline.

\section{Velocity Reconstruction} \label{sec:Reconstruction}

In this section, we present a framework for reconstructing the local DM velocity distribution based on the halo’s accretion history. As outlined above, we decompose the local DM population into four components:
(1) the diffuse component, consisting of smoothly accreted material and unresolved low-mass mergers;
(2) the top four mergers, defined as the most significant contributors to the local stellar population and traceable via stellar kinematics;
(3) young untraceable mergers, which are recent and massive but lack clearly identifiable stellar counterparts; and
(4) old untraceable mergers, which are sufficiently ancient or phase-mixed such that their stellar debris can no longer be reliably associated with a specific progenitor.
This classification forms the foundation of our reconstruction model, in which the full DM velocity distribution is expressed as the weighted sum of each component.

The diffuse component is accreted gradually and, by construction, does not contribute to the local stellar population. As shown in Sec.~\ref{sec:Diffuse}, this component is well phase-mixed by $z=0$ and exhibits velocity distributions that are well described by generalized Gaussian distributions. This allows for a compact parametric description of the diffuse DM kinematics, which we incorporate into the overall reconstruction model. The young and old untraceable mergers, as we discussed before, will also be assumed to follow the same distribution as the diffuse component. So the full velocity distribution can be written as in eq.~\ref{full_distribution}
\begin{equation}
    f_{DM} = \sum_{i=1}^m c_i f_i + (1-\sum_{i=1}^m c_i) f_{\rm{relaxed}}
\end{equation}
where $c_i$ denotes the DM mass fraction of the i-th top merger (with $m=4$ in this work), and $f_i$ their velocity distribution. The remaining mass fraction is assigned to the relaxed component, $f_{\rm{relaxed}}$, which describes both the diffuse material and untraceable mergers. Ideally, all components in this decomposition could be determined from direct observations. In practice, however, none of them can be directly observed so we will have to make the following assumption. The DM velocity distribution $f_i$ for each of the merger can be inferred from the stellar distribution as described in Sec.~\ref{subsec:Matching}, while the remaining material—including the diffuse component and all untraceable mergers—is assumed to follow the generalized Gaussian form characterized in Sec.~\ref{sec:Diffuse}.

At this stage, the only remaining unknown in our reconstruction framework is the mass fraction of DM contributed by each merger to the solar neighborhood. Observationally, it is extremely challenging to directly determine how much DM each progenitor deposits locally. However, since DM and stars are accreted together—though with different spatial distributions and disruption timescales—it is plausible that the DM mass loss fraction relative to the progenitor is correlated with the corresponding stellar mass loss fraction. Moreover, the total stellar mass of each progenitor and its stellar debris are more accessible observationally, making the stellar mass loss fraction a potentially useful proxy for estimating the deposited DM.

\begin{figure}[t]
    \centering
    \includegraphics[width=1\linewidth]{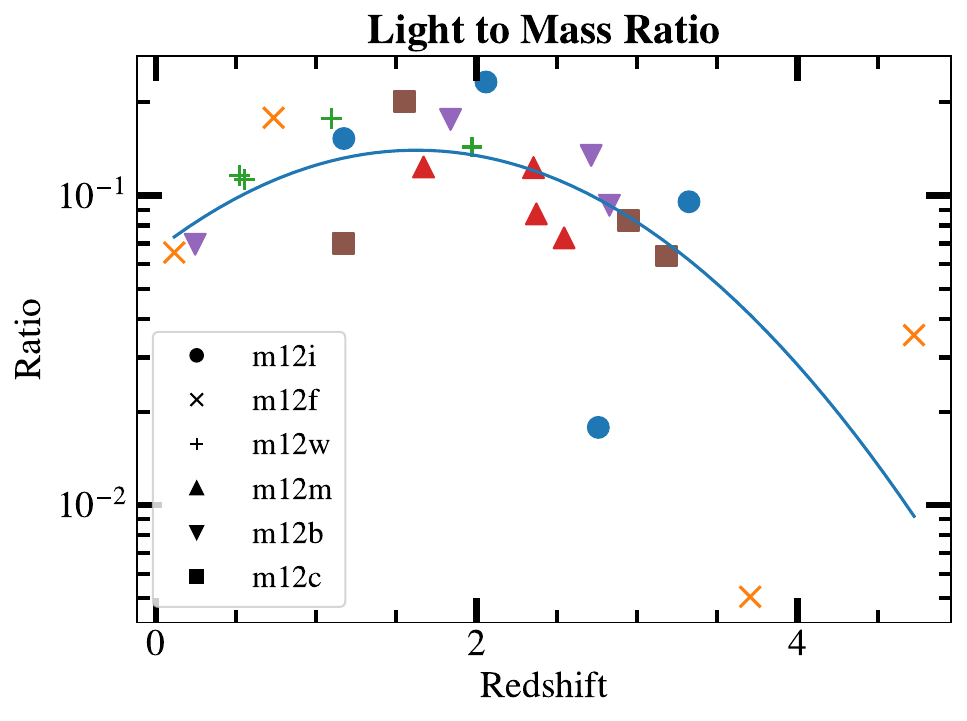}
    \caption{\label{fig:ratio}Ratio of the deposited DM fraction to the deposited stellar fraction in the solar neighborhood, plotted as a function of accretion redshift for the top four mergers across all six \texttt{m12} simulations. The y-axis shows the quantity $r=c_i/c_i^*$, where $c_i$ is the local DM mass fraction and $c_i^*$ is the local stellar mass fraction for each merger. A clear redshift-dependent trend is observed: the ratio increases with redshift up to a turnover point around $z\sim2$, and then declines at earlier times. This behavior reflects the differential deposition and relaxation timescales of DM and stellar material, and is used to relate observable stellar fractions to the corresponding DM contributions in the reconstruction model.}
\end{figure}

To investigate this, we compute the ratio of the DM loss fraction to the stellar loss fraction for each merger (Fig.~\ref{fig:ratio}). The ratio rises with increasing redshift before turning over around $z\sim2$, reflecting the different stripping histories of the two components. Because stars are more tightly bound and centrally concentrated within the progenitor, they are stripped later and deposited at smaller host-centric radii; DM, being more diffusely distributed, is stripped earlier and spread more broadly through the host halo.

For recent mergers (i.e., $z\lesssim2$), DM is still being deposited into the solar neighborhood, so the local DM fraction grows with decreasing redshift. For older mergers ($z\gtrsim2$), most material has already been deposited, but subsequent accretion events inject energy into earlier debris via violent relaxation, preferentially pushing the loosely bound DM to larger orbits. The more tightly bound stellar component is less susceptible to this redistribution, which may explain why the local DM fraction from old mergers declines at early accretion times while the stellar contribution remains comparatively stable.

Consequently, the turnover in the DM-to-stellar loss ratio is largely driven by the evolution of the DM deposition process. Given this behavior, we use the stellar mass loss fraction as a proxy to infer the corresponding DM fraction. We fit the ratio of DM-to-stellar loss as a function of redshift using a parabolic model in log-space, and find that the trend can be well described by
\begin{equation}
    r(z) = \exp(-(0.198 \pm 0.083)z^2+(0.372 \pm 0.53)z-(2.4\pm0.38)))\label{eq:ltm}
\end{equation}
where $r(z) \equiv c_i/c_i^*$ and $c_i^*$ is the deposited stellar mass fraction for the $i$-th merger. The best-fit relation exhibits substantial scatter, and the corresponding uncertainties in the fitted parameters are non-negligible. As a result, the inferred ratio can differ from the true value by up to a factor of a few in some cases. When combined with abundance matching, this step constitutes the dominant source of uncertainty in the reconstruction of the local velocity distribution. Using this relation, we express the fully reconstructed DM velocity distribution for each component, based purely on observable stellar quantities, as
\begin{eqnarray}
    f_{\rm{DM}}(v_k) &= \sum_{i=1}^4 r(z)c_i^* f_i^**N(0, \sigma_k(z)) \nonumber\\  
    &+ (1-\sum_{i=1}^4 r(z)c_i^*) N_g(0, \alpha, \beta)
\end{eqnarray}
with $k \in \{r, \theta, \phi\}$, $f_i^*$ as the stellar velocity distribution of the $i$-th merger, convolved with a Gaussian kernel as described in Sec.~\ref{subsec:Matching}, and $N_g(0, \alpha, \beta)$ denoting the generalized Gaussian profile (defined in Eq.~\ref{eq:generalized_gaussian}) used for the diffuse component as discussed in Sec.~\ref{sec:Diffuse}.

The reconstructed velocity distributions for \texttt{m12i} and \texttt{m12m} are shown in Fig.~\ref{fig:recon}, with the other galaxies shown in Appendix.~\ref{sec:other}. The solid lines represent different velocity components stacked together, while the black dashed lines represent the gSHM. The true distributions is presented as histograms with different component stacked together. 

The results show that the reconstruction reproduces the true dark matter velocity distributions with good overall accuracy. Each curve broadly follows the corresponding simulated distribution for that constituent. In addition, the generalized Gaussian model is able to capture many of the same features, including the lighter tails and broader central structure. One notable exception, however, is the azimuthal component, where the full reconstruction performs significantly better than the generalized model. This is expected, as the SHM assumes velocity isotropy, which is broken in realistic halos due to anisotropic infall and merger-induced angular momentum. In particular, major mergers can drive coherent co-rotation in the stellar and DM components~\citep{2009MNRAS.397...44R, Purcell_2009, Pillepich_2014}, leading to deviations from the SHM prediction. These effects are more accurately captured by incorporating the stellar kinematics of individual mergers into the reconstruction.

\begin{figure*}[t]
    \centering
    \includegraphics[width=1\linewidth]{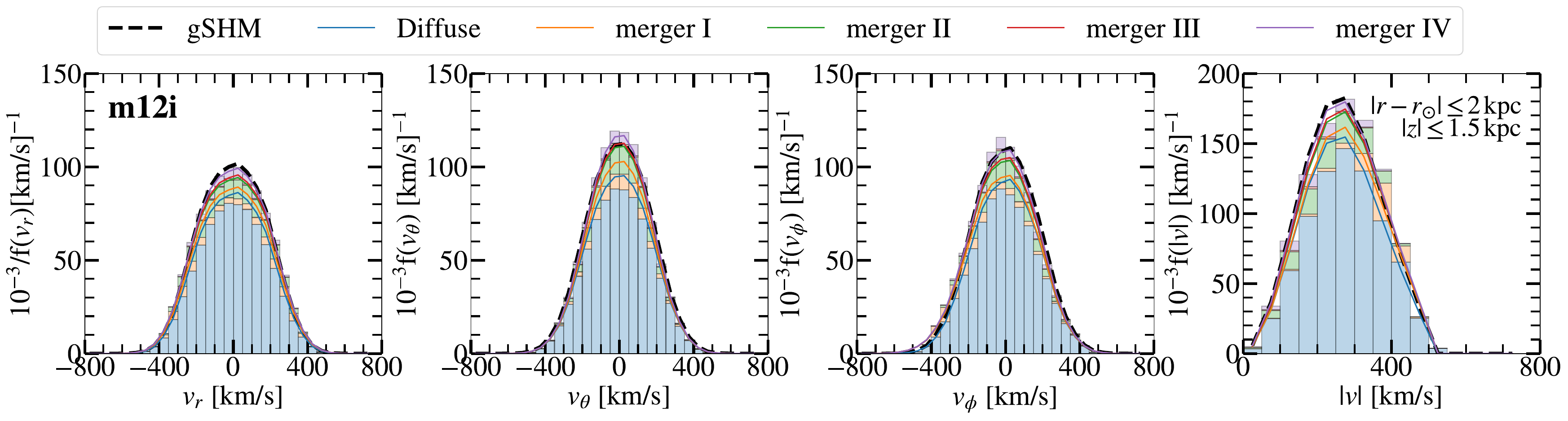}
    \includegraphics[width=1\linewidth]{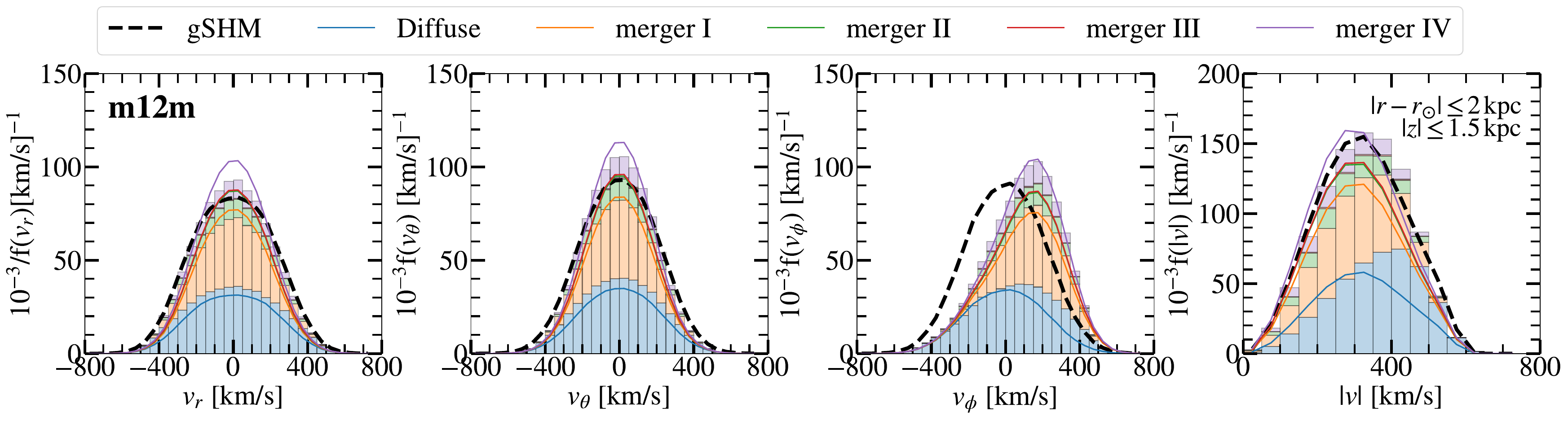}
    \caption{\label{fig:recon} Reconstructed local DM velocity distributions for halos \texttt{m12i}, and \texttt{m12m}. Solid lines represent the reconstructed velocity components in cylindrical coordinates, obtained by combining contributions from the top four mergers (via convolved stellar velocity distributions) and a generalized Gaussian component representing the relaxed material. The blue dashed lines show the true DM velocity distributions from the simulation, while the yellow dotted lines correspond to the predictions from the SHM and generalized SHM. The reconstructions show especially strong improvement in the azimuthal direction, where deviations from SHM arise due to merger-induced co-rotation and anisotropy in the halo.}
\end{figure*}

More precisely, we compute the normalized EMDs between the true DM speed distributions and three models: the SHM, the gSHM, and our full reconstructed model. These comparisons are shown in Fig.~\ref{fig:model}, allowing a direct assessment of how well gSHM and the reconstruction perform relative to the SHM baseline. The results show that both gSHM and the reconstruction consistently outperform the SHM across all halos and velocity components. However, there is no clear preference between gSHM and the reconstruction themselves. This outcome reflects a trade-off; while the reconstruction offers improved modeling of merger-induced DM kinematics by incorporating boosted stellar velocity distributions, it is also more sensitive to uncertainties in the inferred mass fractions. These introduce a source of systematic error, which cannot be resolved without a more accurate understanding of how DM is deposited during tidal disruption.

\section{Discussion} \label{sec:Discussions}

In this section, we discuss the uncertainties and limitations of the method in Sec. ~\ref{subsec:Uncertainties}, and then test across resolutions in Sec. ~\ref{subsec:Resolutions}. 

\subsection{Uncertainties and Limitations}\label{subsec:Uncertainties}

\begin{figure*}[t]
    \centering
    \includegraphics[width=1\linewidth]{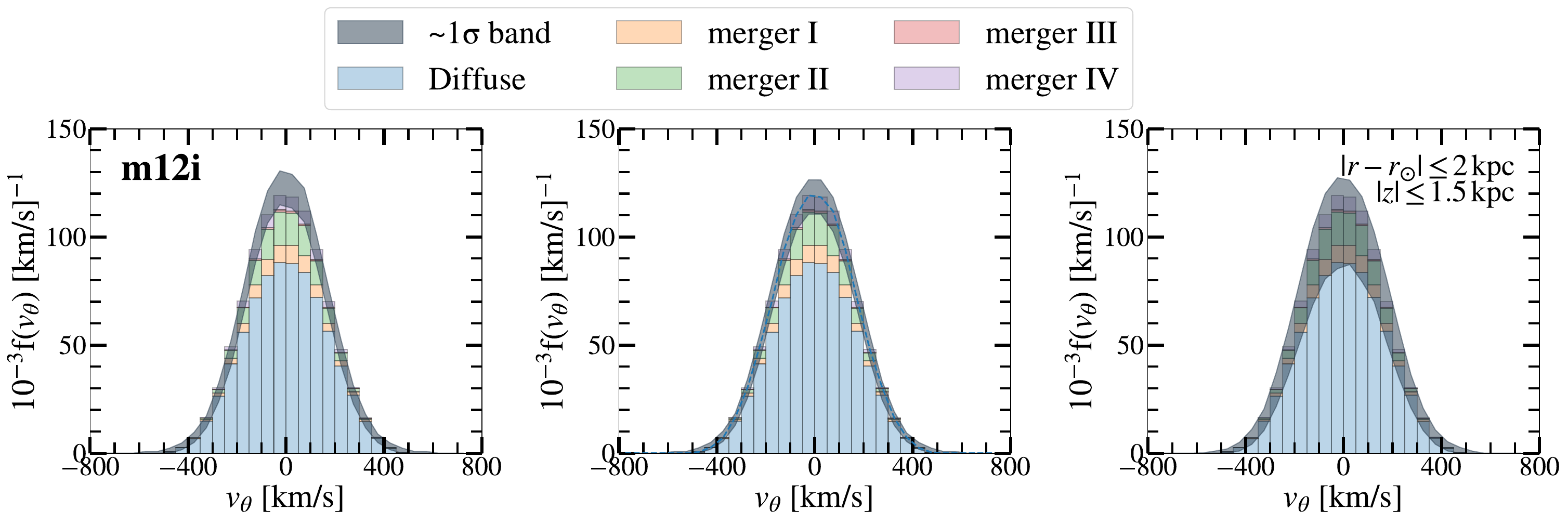}
    \caption{\label{fig:err}Step-by-step illustration of how different modeling assumptions affect the reconstructed DM $v_{\theta}$ velocity distribution for the \texttt{m12i} halo. Each panel shows the stacked velocity distribution across all components, compared to the true DM distribution. Left: Reconstruction using the true DM mass fractions for the top four mergers and the generalized Gaussian for the diffuse component, without convolution. Center: Same as left, but with Gaussian convolution applied to the stellar velocity distributions of the mergers. The agreement improves, with a modest increase in uncertainty. Right: Reconstruction using inferred mass fractions from the light-to-mass ratio model (Eq.~\ref{eq:ltm}), instead of the true values. The discrepancy with the true distribution increases significantly, and the uncertainty grows substantially. Shaded regions represent one-sigma uncertainty bands, and solid lines indicate the reconstructed means. This comparison highlights that mass fraction estimation is the dominant source of uncertainty in the reconstruction framework.}
\end{figure*}

The results presented in the previous section demonstrate that the local DM velocity distribution can be meaningfully reconstructed from a galaxy’s accretion history and its associated stellar debris. By explicitly modeling contributions from both diffuse and merger-driven components, our framework captures key dynamical features that are absent in the SHM. The reconstruction performs particularly well in the azimuthal velocity component, underscoring the importance of merger-induced kinematic substructure in shaping the local DM phase-space distribution. However, despite these improvements, the reconstruction does not exhibit a clear or systematic advantage over the gSHM in terms of speed distribution. As shown in Sec.~\ref{subsec:Matching}, the Gaussian convolution consistently improves the agreement between stellar and DM kinematics for individual mergers. The absence of a comparable improvement at the level of the full reconstruction therefore points to uncertainties associated not with the kinematic matching itself, but with the relative weighting of merger contributions—most notably, the inferred light-to-mass ratios.

To isolate the sources of uncertainty in the reconstruction, Fig.~\ref{fig:err} illustrates the impact of each modeling step for $v_{\theta}$ of \texttt{m12i} as an example. In the leftmost panel, we reconstruct the velocity distribution by stacking the stellar velocity distributions together with the sampled generalized Gaussian for the diffuse component, weighted by the true relative DM mass fractions of the mergers. The shaded region reflects the uncertainty associated with the generalized Gaussian parameters. While the reconstructed mean differs slightly from the true DM distribution, it remains statistically consistent within the uncertainties. In the middle panel, we additionally apply the Gaussian convolution to the merger stellar distributions. This step significantly improves the agreement between the reconstructed and true distributions, although the uncertainty band broadens modestly due to the additional convolution parameters. In contrast, the rightmost panel shows the result when the relative mass fractions are no longer taken from the simulation truth but instead inferred using Eq.~\ref{eq:ltm}. In this case, the reconstructed mean deviates substantially from the true distribution, and the associated uncertainty increases markedly.

Taken together, these results indicate that the dominant limitation of the reconstruction arises from uncertainties in estimating the dark-to-stellar mass ratios of individual mergers. While the kinematic mapping between stars and DM can be robustly modeled, the mass assignment introduces significant systematic uncertainty. This points to an incomplete understanding of how DM is deposited during tidal disruption and violent relaxation. Addressing this limitation is fundamentally a modeling challenge that cannot be resolved through improved observations alone, but instead requires more detailed theoretical and numerical studies of the accretion and relaxation processes governing DM in hierarchical galaxy formation.

\subsection{Resolution Effects}\label{subsec:Resolutions}

\begin{figure*}[t]
    \centering
    \includegraphics[width=1\linewidth]{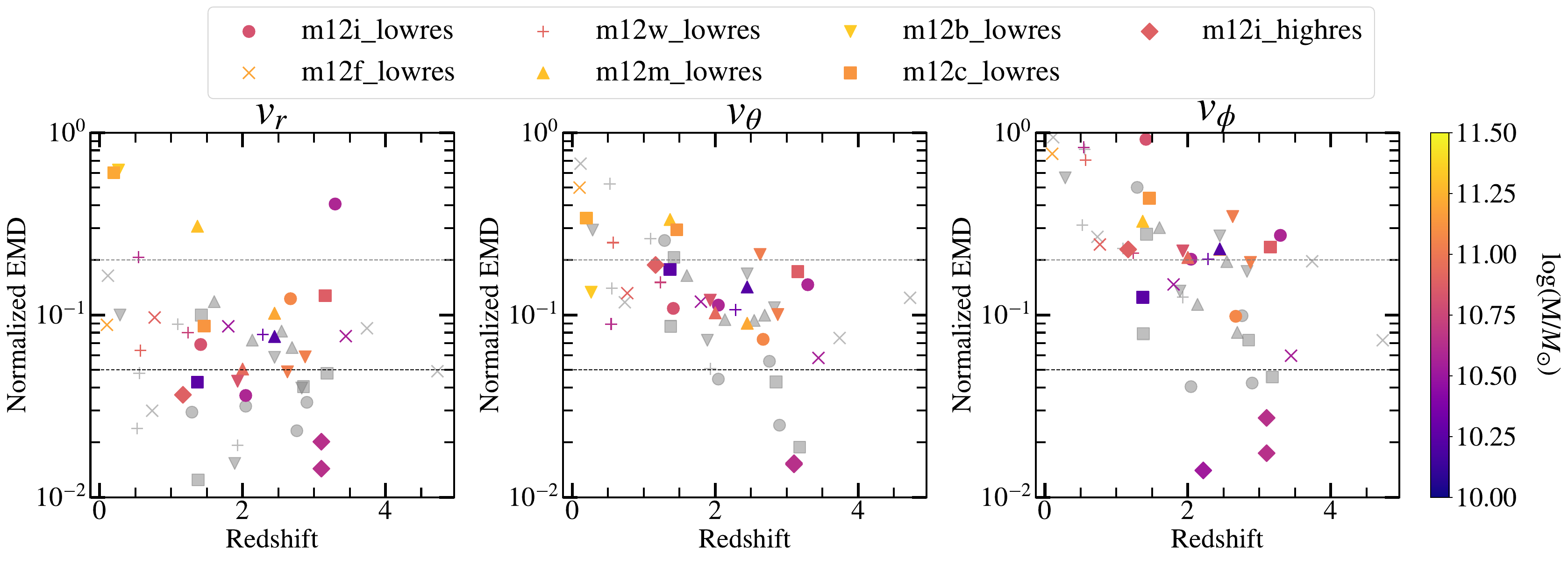}
    \caption{\label{fig:EMD-RTS-res}EMD for $v_r$, $v_t$, and $|\vec{v}|$ between the stellar and DM velocity distributions for mergers across different simulation resolutions. The fiducial resolution is shown in gray. Low-resolution mergers generally exhibit larger EMD values than their higher-resolution counterparts, while following similar redshift-dependent trends. A small number of mergers show exceptionally large EMDs in the radial component, indicating genuine resolution limitations. }
\end{figure*}

\begin{figure}[t]
    \centering
    \includegraphics[width=1\linewidth]{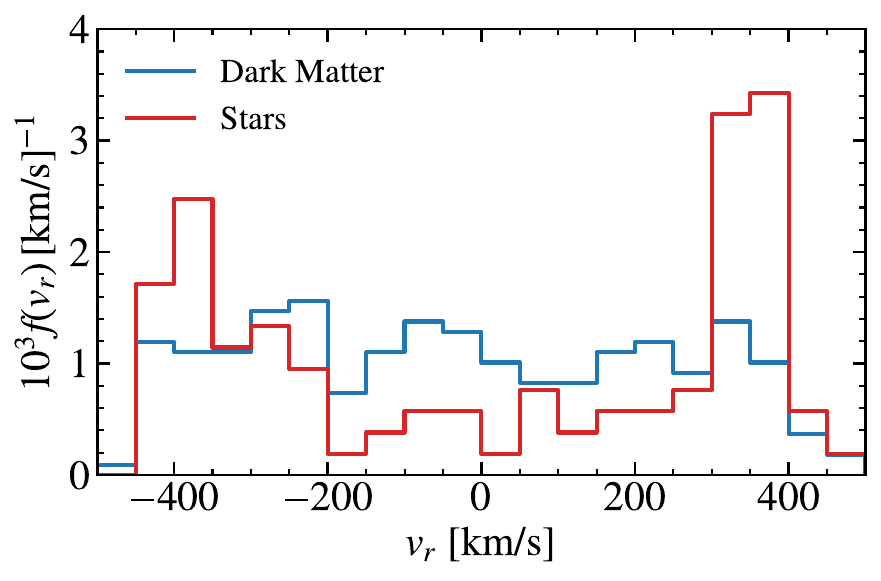}
    \caption{\label{fig:iw-m4-low} Illustration of a resolution-driven failure mode in the EMD analysis for a low-mass merger in the low-resolution \texttt{m12w} run. Shown are the stellar and DM velocity distributions associated with the fourth-ranked merger (by local stellar contribution). Owing to the limited number of star and DM particles in this event, the sampled velocity distributions become irregular and noise-dominated. }
\end{figure}

In this section, we assess the robustness of our analysis across different simulation resolutions. As described in Sec.~\ref{sec:sim}, all of our \texttt{m12} simulations include lower-resolution runs with particle masses eight times larger than those in the standard \textit{Latte} suite. Additionally, we utilize the only higher-resolution realization — the Triple Latte simulation — available for \texttt{m12i} (Wetzel el al. in prep).\footnote{It is important to note that simulations at different resolutions share the same initial conditions, ensuring broadly similar accretion histories. However, due to numerical differences, the precise timing of merger events, and in particular the impact angles and positions, may vary. As a result, direct one-to-one comparisons of velocity distributions between simulations at different resolutions are not meaningful.} 

To identify corresponding mergers across simulations of different resolutions, we match subhalos using their peak masses and broad accretion histories, supplemented by visual inspection of their trajectories through time. We do not rely on the merger indices assigned by the Rockstar halo finder, since these labels are not stable across resolutions. Although the overall subhalo mass hierarchy is usually preserved, the associated stellar content can differ appreciably between runs, largely due to resolution-dependent particle assignment in Rockstar. Consequently, the ordering of mergers by local stellar contribution—particularly beyond the most dominant events—is not always preserved. For example, a merger labeled “4” or “5” in a low-resolution run need not correspond to the same physical progenitor as the similarly labeled merger in the fiducial-resolution simulation.

Repeating our EMD analysis for these alternative-resolution runs yields the results shown in Fig.~\ref{fig:EMD-RTS-res}. We find that the low-resolution simulations systematically exhibit slightly higher EMD values compared to the standard Latte runs, though they generally follow the same redshift trend. Notably, some mergers display significantly elevated EMDs—especially in the radial velocity component—indicative of true resolution limitations. In particular, lower-mass events (often corresponding to the fourth-ranked merger by local stellar contribution) can be represented by too few stellar and dark-matter particles for their velocity distributions to be robustly sampled. This undersampling can produce irregular, noise-dominated velocity histograms and consequently unreliable EMD estimates, as illustrated in Fig.~\ref{fig:iw-m4-low}.

This observation is especially relevant when comparing our results to other simulation suites, such as TNG50, as analyzed in \citep{shpigel2025buildempiricalspeeddistribution}. In that work, the highest DM mass resolution is $3.07\times10^5 M_{\odot}/h\approx4.5\times10^5 M_{\odot}$ for $h=0.6774$ in TNG50-1~\cite{Nelson_2019, Pillepich_2019}, comparable to the particle mass in our low-resolution runs. Consequently, the larger EMD values reported for some mergers in the TNG simulations may partly reflect resolution limitations rather than intrinsic differences in the accretion or kinematic structure. In contrast, our higher-resolution Triple Latte realization for \texttt{m12i} yields results consistent with the standard Latte analysis, suggesting that our primary conclusions are not strongly affected by numerical resolution. This highlights the importance of accounting for resolution effects when interpreting discrepancies across different simulation frameworks.

For the diffuse component, we assess the stability of its mass fraction across resolution levels which shows broadly consistent behavior with the results in Fig.~\ref{fig:diffuse_fraction}. We then perform fits to the velocity distributions using the generalized Gaussian profile, as done previously. The generalized Gaussian continues to provide better fit than the standard Gaussian regardless of resolution, confirming its robustness. The best-fit parameters for the low- and high-resolution simulations are also presented in Table.~\ref{tab:best_fit}, all of which also remain consistent with our fiducial results. It is important to note that for the standard- and low-resolution simulations, the quoted uncertainties include both the fitting errors and the statistical variance across the six \texttt{m12} runs. In contrast, the high-resolution result is derived from \texttt{m12i} alone, so the uncertainties reflect only the fitting error.

These findings confirm that the phase-space structure of the diffuse DM component—and the associated generalized Gaussian description—are robust against changes in numerical resolution.

\section{Conclusions} \label{sec:Conclusion}

In this work, we present a framework for reconstructing the local DM velocity distribution by leveraging a galaxy’s accretion history, guided by stellar kinematics and insights from high-resolution cosmological simulations. We decompose the DM population into four distinct components—diffuse DM, top mergers traceable by stellar debris, and both young and old untraceable mergers—and construct a composite model that integrates both parametric descriptions and data-driven elements to capture the full kinematic complexity of the halo.

We show that the diffuse DM component is well characterized by generalized Gaussian profiles, while the contributions from major mergers can be effectively modeled by convolving stellar velocity distributions with Gaussian kernels. Using the EMD as a quantitative diagnostic, we find that our reconstruction consistently outperforms the Standard Halo Model (SHM), and that the generalized SHM (gSHM) alone also represents a significant improvement over SHM. However, due to current limitations in modeling the DM mass contribution from individual mergers—primarily stemming from incomplete understanding of the accretion and tidal disruption processes—the full reconstruction does not exhibit a systematic advantage over gSHM. Consequently, we recommend the generalized SHM as the most robust and practical refinement of the SHM for use in future DM direct-detection analyses. For directional signals and annual modulation probes, we recommend the full reconstruction.

This approach highlights the promise of using stellar information as a proxy for the local dark matter structure, while also making clear that the dominant uncertainty lies in estimating the amount of dark matter contributed by each merger. In practice, this requires inferring the accreted dark matter fraction from the stellar component through the light-to-mass ratio and abundance matching, both of which remain intrinsically uncertain and are not expected to improve dramatically in the near term. As a result, our reconstruction is already close to the best that can presently be achieved within this framework. Meaningful further progress will likely require a better theoretical and numerical understanding of tidal disruption, mass deposition, and the relation between stellar and dark matter stripping in hierarchical galaxy formation.

Furthermore, the best-fit $\beta$ values in the gSHM are consistently similar across the angular directions, while the radial direction exhibits a systematically higher $\beta$, indicating a more boxy distribution with suppressed tails. This anisotropy may reflect the effects of violent relaxation along the radial direction, hinting at a statistical mechanics origin for the generalized Gaussian form. The consistency of these features across different simulations and resolutions points toward an emergent dynamical equilibrium. Further theoretical investigations are warranted to explore the origin and generality of this behavior.

The reconstruction pipeline as well as the generalized SHM can be straightforwardly applied to observational data and thus modify the implied MW local DM velocity distribution. The implications of these new velocity distributions on DM direct detection experiments will be explored in subsequent works.

\begin{acknowledgments}
We thank E. Y. Davies, D. Folsom, A. Hussein, M. Lisanti, Z. Mezghanni, T. Schpigel, N. Starkman for helpful conversations. 
XZ, AT, and LN received support from NSF, via CAREER award AST-2337864.
L.N. is supported by the Sloan Fellowship and the NSF CAREER award AST-2337864.
AW received support from NSF, via CAREER award AST-2045928. 
AA acknowledges support from Gordon and Betty Moore foundation.

FIRE-2 simulations are publicly available \citep{Wetzel_2023, wetzel2025secondpublicdatarelease} at \url{http://flathub.flatironinstitute.org/fire}.
Additional FIRE simulation data is available at \url{https://fire.northwestern.edu/data}.
A public version of the \textsc{Gizmo} code is available at \url{http://www.tapir.caltech.edu/~phopkins/Site/GIZMO.html}.
We generated simulations using: XSEDE, supported by NSF grant ACI-1548562; Blue Waters, supported by the NSF; Frontera allocations AST21010 and AST20016, supported by the NSF and TACC; Pleiades, via the NASA HEC program through the NAS Division at Ames Research Center. We acknowledge the Texas Advanced Computing Center (TACC) at The University of Texas at Austin for providing computing resources that contributed to our results.

\end{acknowledgments}

\begin{contribution}
XZ led the analysis, developed the methodology, wrote the analysis code, and drafted the manuscript. AT implemented and debugged the stellar and dark matter assignment pipeline, carried out validation tests, and wrote the appendices and associated sections. LN conceived the project, provided the initial codebase, and supervised the analysis throughout. AW and AA provided feedback on the methodology and results within the framework of the FIRE collaboration. All authors reviewed and commented on the manuscript.
\end{contribution}

\software{
NumPy~\citep{Harris2020NumPy}, 
SciPy~\citep{Virtanen2020SciPy}, 
Matplotlib~\citep{Hunter2007matplotlib},
Pandas~\citep{mckinney2010pandas, pandas_software}, 
Astropy~\citep{2013A&A...558A..33A,2018AJ....156..123A,2022ApJ...935..167A}, 
GizmoAnalysis~\citep{2020ascl.soft02015W}, 
HaloAnalysis~\citep{2020ascl.soft02014W}, 
Rockstar~\citep{Behroozi_2012a},
Consistent Trees~\citep{2012ascl.soft10011B}
          }

\pagebreak
\newpage

\clearpage 

\setcounter{page}{1}
\setcounter{table}{0}
\setcounter{figure}{0}
\setcounter{equation}{0}
\renewcommand{\thepage}{\roman{page}}
\renewcommand{\thefigure}{S\arabic{figure}}
\renewcommand{\thetable}{S\arabic{table}}
\renewcommand{\theequation}{\thesubsection.\arabic{equation}}

\appendix

\section{GSE-like Mergers}\label{subsec:GSE}

The GSE is one of the most consequential merger events in the Milky Way’s history~\citep{2018MNRAS.478..611B,2018Natur.563...85H}. It is believed to have merged with the Milky Way approximately 8–11 Gyr ago (corresponding to redshift $z \sim 1$-2) with an estimated peak mass of $\sim10^{11}M_{\odot}$~\citep{johnson2025thatsretrogaiasausageenceladusmerger}, depositing a substantial fraction of the inner stellar halo and heating the early Galactic disk~\citep{2018Natur.563...85H, 2020A&A...642L..18K}. It left behind a dynamically hot, metal-poor stellar population with distinct radial kinematics. 

\begin{figure}[h]
    \centering
    \includegraphics[width=0.5\linewidth]{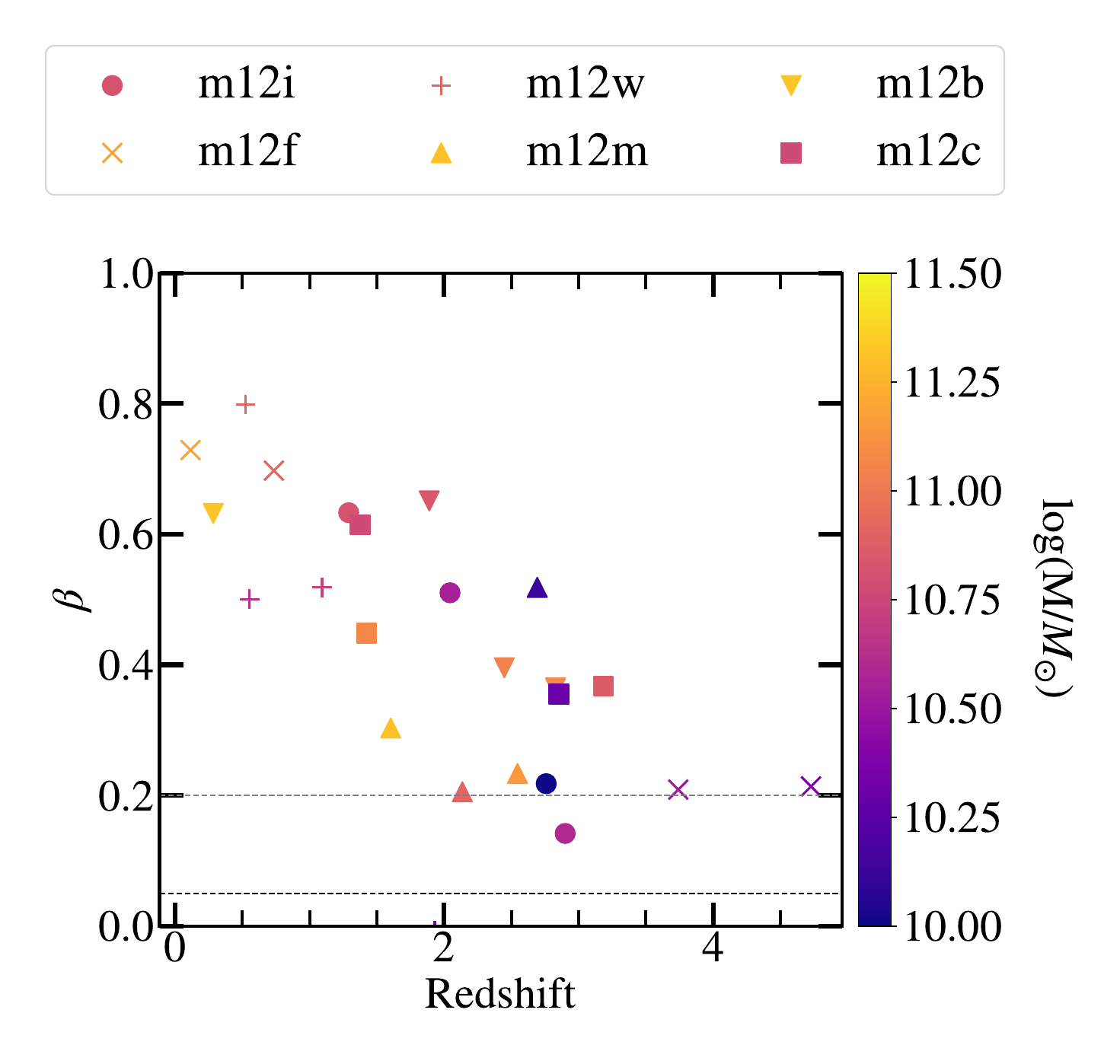}
    \caption{\label{beta}Velocity anisotropy parameter $\beta$ at $z=0$ as a function of accretion redshift for individual mergers across all \texttt{m12} galaxies. Each point represents a merger event, with $\beta$ computed from the present-day kinematics of its associated DM particles.}
\end{figure}

A useful way to characterize the orbital nature of a merger is through the velocity anisotropy parameter $\beta$, defined as
\begin{equation}
\beta = 1-\frac{\sigma_{\theta}^2+\sigma_{\phi}^2}{2\sigma_r^2},\label{eq:beta}
\end{equation}
where $\sigma_{r, \theta, \phi}$ are the velocity dispersions for the corresponding component. GSE debris today is observed to have a highly radial velocity distribution, with $\beta \sim 0.8$–0.9, indicative of an eccentric, plunging orbit~\citep{2018MNRAS.478..611B, Lane_2021}. The velocity anisotropies at $z=0$ of individual mergers in our sample are shown in Figure~\ref{beta}. 

Based on these properties, we define “GSE-like” mergers in our simulations as those that satisfy:

1. Moderate-to-high mass: $M>5\times 10^{10}M_{\odot}$

2. Radial orbit: $\beta>0.6$

3. Intermediate accretion time: $0.5 < z < 1.5$

\section{Limitations on merger history}\label{sec:limitation}

In this appendix, we discuss limitations in the reconstruction of merger histories and the steps taken to mitigate them. The limitations include misassociation originated from flyby events and subhalo-subhalo mergers that potentially would be misidentified as separate merger events. In the following subsections, we discuss each effect and the consistency checks used to identify and mitigate it. 

\subsection{Flyby rate}\label{sec:flyby}

One such limitation arises from the flyby rate, in which DM particles that are not genuinely associated with an accreting subhalo are nevertheless identified as part of a merger by the algorithm. This effect is negligible for star particles, which are preferentially concentrated near the centers of subhalos and therefore less susceptible to misclassification. In this work, we define a flyby particle as a particle already within the host halo radius prior to formal accretion of the subhalo. The original convention was to trace particles backward in time, requiring association in three of the last four snapshots, with a velocity dispersion cut of 3$\sigma$.

We found that the flyby rate is dependent on the mass of the oncoming subhalo. Larger subhalos attracted a greater number of flybys from the main halo than subhalos with smaller mass. Table \ref{tab:contamination_by_mass} displays representative mergers from \texttt{m12i} and \texttt{m12f} and their corresponding contamination rates.

\begin{table*}[t]
\centering
\caption{Representative mergers from \texttt{m12i} and \texttt{m12f} illustrating the dependence of flyby contamination on subhalo peak mass. 
The contamination rate is defined as the fraction of particles associated with a merger that are located within the host halo at a snapshot preceding accretion. 
Higher-mass subhalos exhibit larger flyby contamination fractions.}
\label{tab:contamination_by_mass}
\begin{tabular}{lcccc}
\hline\hline
 & \multicolumn{2}{c}{\textbf{m12i}} & \multicolumn{2}{c}{\textbf{m12f}} \\
\cline{2-3}\cline{4-5}
 & $M_{\rm peak}\,[M_\odot]$ & Contamination (\%) 
 & $M_{\rm peak}\,[M_\odot]$ & Contamination (\%) \\
\hline
\textbf{Merger I}   & $6.45\times10^{10}$ & 3.2 
                    & $1.53\times10^{11}$ & 4.1 \\

\textbf{Merger II}  & $3.55\times10^{10}$ & 0.3 
                    & $8.20\times10^{10}$ & 0.4 \\

\hline
\end{tabular}
\end{table*}

To reduce contamination from flyby DM particles, we imposed a stricter accretion criterion. Specifically, a DM particle was required to be associated with a subhalo in at least six of the last nine snapshots, rather than three of the last four. This modification significantly reduced the number of DM particles misassigned while preserving those truly associated with the merger. Nevertheless, a residual population of flyby particles that were never physically bound to the merger persisted.

\begin{figure*}[t]
    \centering
    \includegraphics[width=0.5\linewidth]{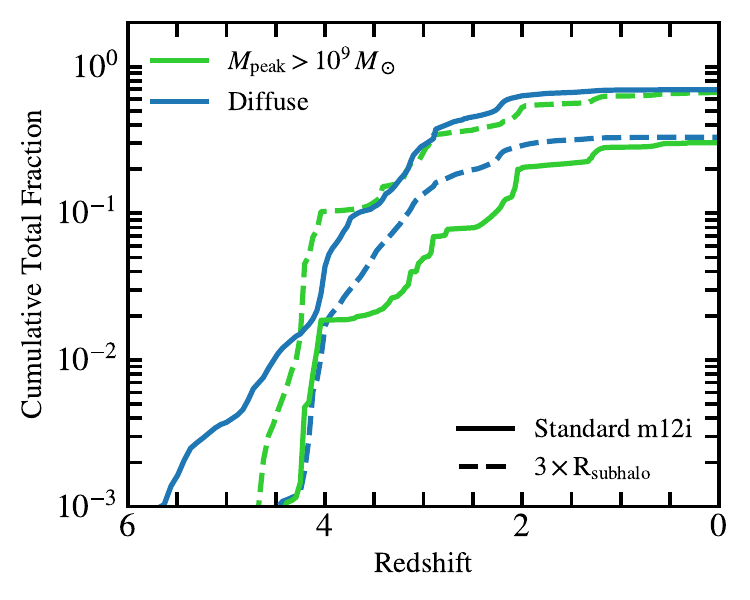}
    \caption{\label{fig:i-subtest} Accretion history of the DM component for the \texttt{m12i} subhalo test, in which the radii of all subhalos (excluding the main halo) were increased by a factor of three, in comparison to it unchanged. The pronounced increase in the diffuse component at $z \sim 4$ for both demonstrates that the diffuse buildup during major merger events persists even under enlarged subhalo radii. This behavior reflects the rapid growth of the main halo during these mergers.}
\end{figure*}

To test the robustness of the merger reconstruction, we performed several consistency checks. First, we increased the radii of all subhalos (excluding the main halo) by factors of 2, 3, and 4 for \texttt{m12i} and \texttt{m12f}. In each case, the number of particles classified as diffuse decreased, as expected. However, pronounced increases in the diffuse component persisted whenever a massive halo was accreted. We attribute these features to the rapid growth of the main halo's radius during such events, leading the main halo to expand and include nearby DM particles that are not physically bound to the merger. 

Fig. \ref{fig:i-subtest} shows the DM merger history for \texttt{m12i} when the subhalo radii (excluding the main halo) are increased by a factor of three. As in Fig. \ref{fig:m12i_merger_history}, the diffuse component exhibits sharp increases at major merger events, reflecting the expansion of the main halo during these periods.

\begin{figure*}[t]
    \centering
    \includegraphics[width=1\linewidth]{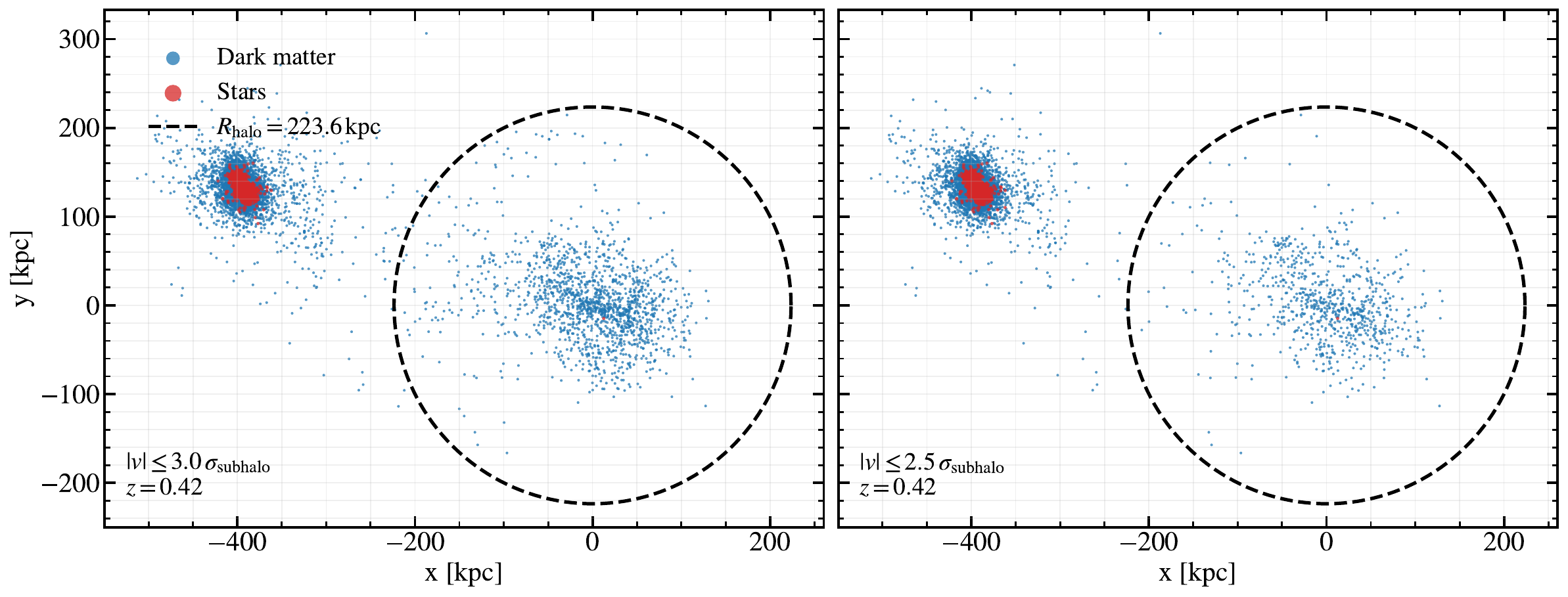}
    \caption{\label{fig:f-com-merger1}Spatial distribution of merger-associated DM (blue) and stellar (red) particles for the same accretion event in the \texttt{m12f} simulation. The left panel shows the original realization, and the right panel shows the corrected realization. Particle positions are shown in the galactic $x$--$y$ plane at a snapshot preceding accretion. The dashed circle marks the main host halo radius at that snapshot. The bottom-left annotation in each panel indicates the dispersion cut $\sigma$ used for merger identification and the corresponding redshift observed before accretion. Differences between panels isolate the impact of the dispersion treatment on the spatial coherence of merger debris.}
\end{figure*}

As a cross-check, we also examined the effect of relaxing the dispersion cut from $2.5\sigma$ to $3\sigma$. As anticipated, this change resulted in a higher flyby rate, confirming that looser kinematic cuts lead to increased contamination from non-merging material. Fig.~\ref{fig:f-com-merger1} demonstrates the increase in contamination rate as we increase $\sigma$ from 2.5 to 3 for Merger 1 of \texttt{m12f}. With $\sigma=2.5$, 4.1\% of the particles associated with Merger 1 are misassociated with that merger. By contrast, this value is 7.6\% for $\sigma=3$.  

Conversely, adopting a stricter criterion reduces flyby contamination but may exclude a small number of genuinely associated merger particles. However, we find that this trade-off is strongly asymmetric: the number of flyby particles introduced when increasing to 3$\sigma$ substantially exceeds the number of genuine merger particles lost when tightening to 2.5$\sigma$. For Merger 1 of \texttt{m12f}, decreasing the cut from 3$\sigma$ to 2.5$\sigma$ removes roughly 600 flyby particles while excluding only about 30 genuinely associated particles. This demonstrates that the 2.5$\sigma$ threshold effectively suppresses contamination while minimally impacting the true merger signal.

Based on the quantitative comparisons above, the final adopted convention for this work is the following: six of nine snapshot requirement, 2.5$\sigma$ dispersion cut, and default subhalo radii.

\subsection{Subhalo-subhalo mergers}

In several cases, mergers identified by the algorithm were required to be combined. This arises because the merger-matching procedure is based primarily on star particle membership. In some situations, a subhalo contained star particles but no associated dark matter particles, leading to the artificial separation of what was physically a single merger event.

\begin{figure*}[t]
    \centering
    \includegraphics[width=0.7\linewidth]{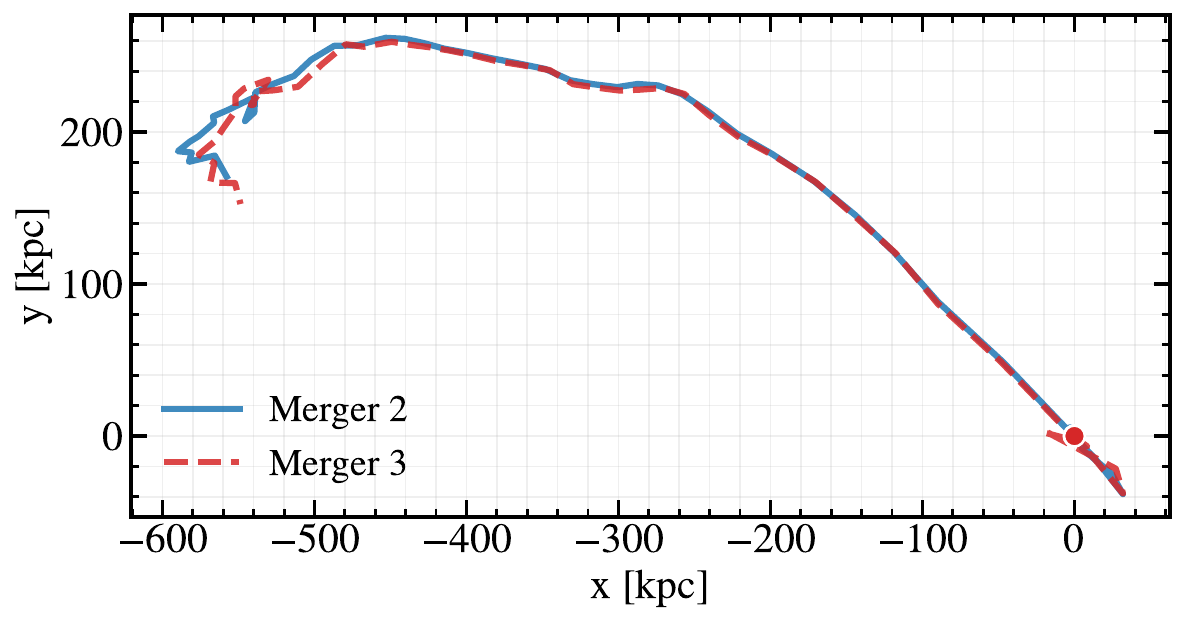}
    \caption{\label{fig:f-2-3-trajectory}Mean stellar orbital trajectories in the $x$--$y$ plane for Mergers 2 and 3 in \texttt{m12f}, traced backward from $z=0$ to $z=3$ with positions relative to the main halo. The point marks the present-day ($z=0$) position of each stellar component. Although initially identified as separate mergers, the stellar particles associated with both mergers follow nearly identical orbital paths at all times, indicating that they originate from the same accreting system.}
\end{figure*}

One example of this behavior occurs for Mergers 2 and 3 in \texttt{m12f}. Merger 2 contained both stellar and DM components, whereas Merger 3 contained a substantial number of star particles but no DM particles. Further inspection revealed that the particles associated with both mergers followed the same orbital trajectory, indicating that they originated from the same accreting system. Consequently, the particles from Mergers 2 and 3 were combined into a single merger, labeled Merger 2. All subsequent merger indices were then shifted accordingly (e.g., Merger 4 $\rightarrow$ Merger 3). Figure \ref{fig:f-2-3-trajectory} displays the trajectories of the two subhalos up to redshift 3, showing how they are from the same system.

The following mergers were combined in this manner:
\begin{itemize}
    \item \texttt{m12f}: Mergers 2 and 3
    \item \texttt{m12m}: Mergers 2 and 5
    \item \texttt{m12i-lowres}: Mergers 3 and 5
\end{itemize}

In other cases, a merger identified via stellar particles had no DM particles assigned to it through our post-processing phase of assigning DM particles (for more details, see \cite{Necib_2019}). Upon further inspection, we found that the DM from the same physical progenitor had been assigned to a separate dark subhalo (one lacking stars). By matching the orbital trajectories of these dark subhalos to the stellar mergers, we were able to reunite the two components.

The following mergers were connected to dark subhalos in this way:
\begin{itemize}
    \item \texttt{m12b}: Merger 4
    \item \texttt{m12c}: Mergers 4 and 5
    \item \texttt{m12i-lowres}: Merger 4
    \item \texttt{m12i-hires}: Merger 4
\end{itemize}

\section{Reconstruction of the Milky Way-like galaxies}\label{sec:other}
\begin{figure*}[h!]
    \centering
    \includegraphics[width=0.8\linewidth]{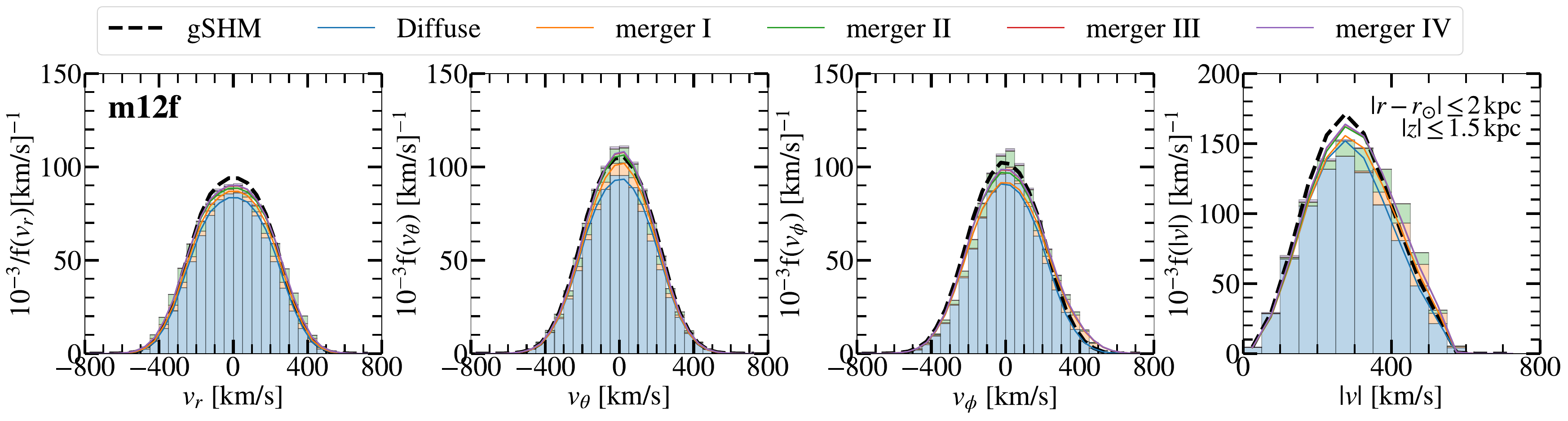}
    \includegraphics[width=0.8\linewidth]{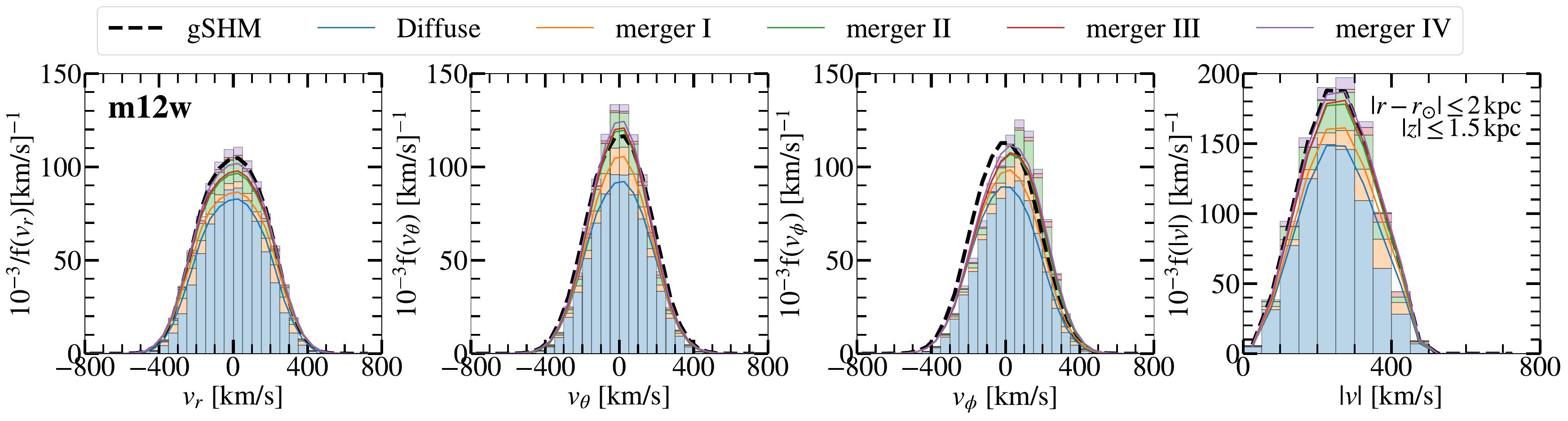}
    \includegraphics[width=0.8\linewidth]{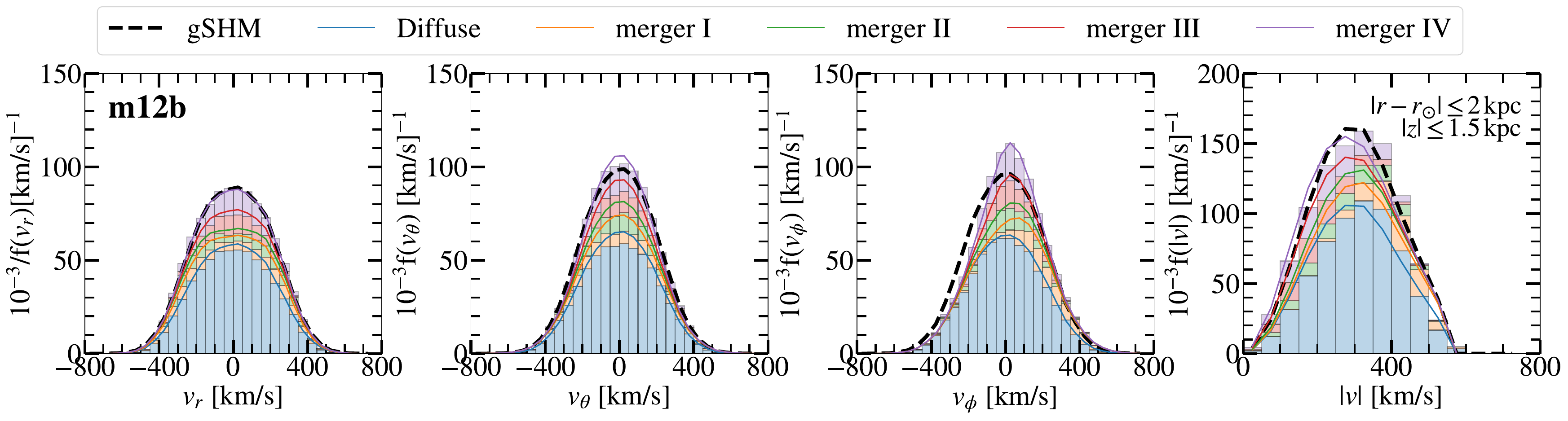}
    \includegraphics[width=0.8\linewidth]{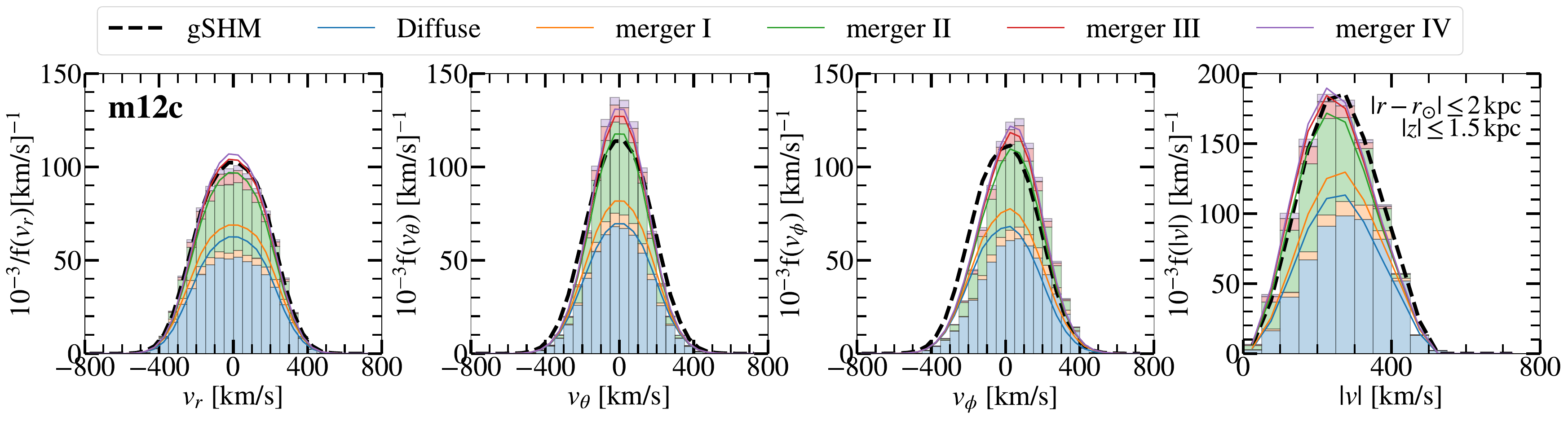}
    \caption{\label{fig:recon_all} Reconstructed local DM velocity distributions for halos \texttt{m12f}, \texttt{m12w}, \texttt{m12b} and \texttt{m12c}. Solid lines represent the reconstructed velocity components in cylindrical coordinates, obtained by combining contributions from the top four mergers (via convolved stellar velocity distributions) and a generalized Gaussian component representing the relaxed material. The blue dashed lines show the true DM velocity distributions from the simulation, while the yellow dotted lines correspond to the predictions from the SHM and generalized SHM.}
\end{figure*}

In this Appendix, we present supplementary results for the \texttt{m12} halos that are not shown in the main text. The exact mass fraction of DM components within the solar neighborhood is presented in Table~\ref{tab:mergerfractionnew}. Merger properties analogous to those in Table~\ref{tab:m12_mergers} are provided for \texttt{m12c}, \texttt{m12m}, \texttt{m12w}, and \texttt{m12b} in Tables~\ref{tab:m12c_m12m_mergers}–\ref{tab:m12w_m12b_mergers}. In addition, the full velocity-distribution reconstructions—analogous to Fig.~\ref{fig:recon}—for the remaining \texttt{m12} halos are shown in Fig.~\ref{fig:recon_all}.

\begin{table*}[t]
\centering
\caption{\label{tab:mergerfractionnew}Table of mass fractions of DM components within the solar neighborhood. The diffuse component constitutes a significant portion of the total local DM, while the contribution from the top mergers varies across simulations. Untraceable mergers typically contribute slightly less than $10\%$ of the total.}
\setlength{\tabcolsep}{12pt}  
\begin{tabular}{lcccccc}
\hline\hline
 & \textbf{\texttt{m12i}} & \textbf{\texttt{m12f}} & \textbf{\texttt{m12m}} & \textbf{\texttt{m12w}} & \textbf{\texttt{m12b}} & \textbf{\texttt{m12c}} \\
\hline
Diffuse & 0.70 & 0.78 & 0.38 & 0.65 & 0.57 & 0.49 \\
Merger 1 & 0.06 & 0.05 & 0.35 & 0.09 & 0.09 & 0.05 \\
Merger 2 & 0.11 & 0.08 & 0.09 & 0.12 & 0.08 & 0.32 \\
Merger 3 & $\lesssim$0.01 & $\ll$0.01 & $\lesssim$0.01 & 0.02 & 0.08 & 0.05 \\
Merger 4 & 0.04 & $\lesssim$0.01 & 0.09 & 0.03 & 0.13 & 0.02 \\
Young untraceable mergers & 0.06 & 0.06 & 0.07 & 0.05 & 0.05 & 0.05 \\
Old untraceable mergers & 0.03 & 0.03 & 0.01 & 0.04 & $\lesssim$0.01 & 0.01 \\
\hline
\end{tabular}
\end{table*}

\begin{table*} [t]
\centering
\begin{tabular}{lcccccccc}
\hline\hline
 & \multicolumn{4}{c}{\textbf{\texttt{m12c}}} & \multicolumn{4}{c}{\textbf{\texttt{m12m}}} \\
\cline{2-5}\cline{6-9}
 & \textbf{I} & \textbf{II} & \textbf{III} & \textbf{IV} & \textbf{I} & \textbf{II} & \textbf{III} & \textbf{IV} \\
\hline
$M_{\rm peak}$ [$M_\odot$] 
& $5.86\times10^{10}$ & $1.18\times10^{11}$ & $7.20\times10^{10}$ & $2.00\times10^{10}$ 
& $2.02\times10^{11}$ & $8.02\times10^{10}$ & $1.35\times10^{10}$ & $1.38\times10^{11}$ \\

$\langle{\rm [Fe/H]}\rangle$ (solar circle) 
& $-1.27$ & $-1.25$ & $-1.82$ & $-1.86$ 
& $-1.29$ & $-1.65$ & $-1.87$ & $-1.59$ \\

$M_{\rm peak}/M_{\star,\rm total}$ 
& $63.2$ & $381$ & $275$ & $244$ 
& $75.4$ & $106$ & $201$ & $132$ \\

$M_{\star,\rm total}$ [$M_\odot$] 
& $9.28\times10^{8}$ & $3.09\times10^{8}$ & $2.62\times10^{8}$ & $8.23\times10^{7}$ 
& $2.68\times10^{9}$ & $7.55\times10^{8}$ & $6.74\times10^{7}$ & $1.05\times10^{9}$ \\

$f_\star$ (solar circle) 
& $0.244$ & $0.170$ & $0.072$ & $0.026$ 
& $0.676$ & $0.127$ & $0.007$ & $0.164$ \\

$f_{\rm DM}$ (solar circle) 
& $0.094$ & $0.431$ & $0.103$ & $0.047$ 
& $0.570$ & $0.150$ & $0.010$ & $0.139$ \\

$z_{\rm acc}$ (mode) 
& $1.17$ & $1.42$ & $3.18$ & $2.95$ 
& $1.67$ & $2.35$ & $2.37$ & $2.55$ \\

\hline
\end{tabular}

\caption{\label{tab:m12c_m12m_mergers}
Properties of the most significant accreted mergers in \texttt{m12c} and \texttt{m12m}. 
For each merger, we list the peak subhalo mass, the mean stellar
metallicity of accreted stars currently in the solar circle, the peak halo-to-stellar mass ratio, the total stellar
mass of the satellite, the stellar and DM mass fractions contributed within the solar circle (with respect
to the total accreted material from subhalos with $M_{\rm peak} > 10^{9}\,M_\odot$), and the most common accretion
redshift of stripped stars.
}
\end{table*}

\begin{table*} [h!]
\centering
\begin{tabular}{lcccccccc}
\hline\hline
 & \multicolumn{4}{c}{\textbf{\texttt{m12w}}} & \multicolumn{4}{c}{\textbf{\texttt{m12b}}} \\
\cline{2-5}\cline{6-9}
 & \textbf{I} & \textbf{II} & \textbf{III} & \textbf{IV} & \textbf{I} & \textbf{II} & \textbf{III} & \textbf{IV} \\
\hline
$M_{\rm peak}$ [$M_\odot$] 
& $8.10\times10^{10}$ & $5.39\times10^{10}$ & $4.27\times10^{10}$ & $2.10\times10^{10}$ 
& $2.08\times10^{11}$ & $6.94\times10^{10}$ & $1.18\times10^{11}$ & $1.10\times10^{11}$ \\

$\langle{\rm [Fe/H]}\rangle$ (solar circle) 
& $-0.94$ & $-1.41$ & $-1.32$ & $-1.70$ 
& $-0.57$ & $-1.44$ & $-1.40$ & $-1.41$ \\

$M_{\rm peak}/M_{\star,\rm total}$ 
& $37.7$ & $95.6$ & $98.2$ & $161$ 
& $24.4$ & $64.1$ & $129$ & $148$ \\

$M_{\star,\rm total}$ [$M_\odot$] 
& $2.15\times10^{9}$ & $5.64\times10^{8}$ & $4.35\times10^{8}$ & $1.30\times10^{8}$ 
& $8.54\times10^{9}$ & $1.08\times10^{9}$ & $9.14\times10^{8}$ & $7.44\times10^{8}$ \\

$f_\star$ (solar circle) 
& $0.616$ & $0.212$ & $0.049$ & $0.045$ 
& $0.720$ & $0.087$ & $0.085$ & $0.009$ \\

$f_{\rm DM}$ (solar circle) 
& $0.252$ & $0.344$ & $0.054$ & $0.098$ 
& $0.214$ & $0.178$ & $0.185$ & $0.215$ \\

$z_{\rm acc}$ (mode) 
& $0.52$ & $1.09$ & $0.55$ & $1.97$ 
& $0.25$ & $1.84$ & $2.83$ & $2.47$ \\

\hline
\end{tabular}

\caption{\label{tab:m12w_m12b_mergers}
Properties of the most significant accreted mergers in \texttt{m12w} and \texttt{m12b}.
For each merger, we list the peak subhalo mass, the mean stellar
metallicity of accreted stars currently in the solar circle, the peak halo-to-stellar mass ratio, the total stellar
mass of the satellite, the stellar and DM mass fractions contributed within the solar circle (with respect
to the total accreted material from subhalos with $M_{\rm peak} > 10^{9}\,M_\odot$), and the most common accretion
redshift of stripped stars.
}
\end{table*}

\begin{figure}[h!]
    \centering
    \includegraphics[width=0.5\linewidth]{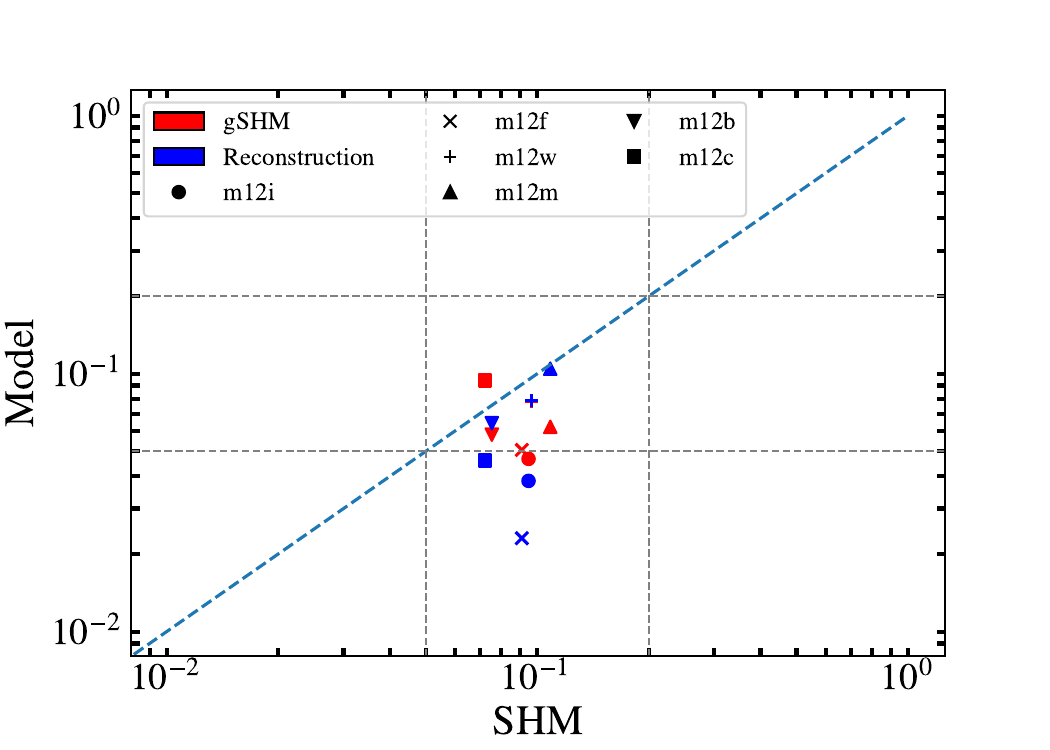}
    \caption{\label{fig:model}Normalized Earth Mover’s Distances (EMDs) between the true DM velocity distributions and three models: the SHM, the gSHM, and the full reconstructed model. The x-axis represents the normalized EMD between the full DM distribution and the SHM, and the y-axis represents the normalized EMD between the full DM distribution and models. }
\end{figure}

\pagebreak
\newpage
\pagebreak

\nolinenumbers
\bibliography{sample7}{}
\bibliographystyle{aasjournalv7}

\end{document}